\documentstyle[12pt,epsfig]{article}
\def\bq{\begin{quote}} 
\def\eq{\end{quote}}

\def\bq{\begin{quote}} 

\def\bar{\overline}

\def\k{m_{eff}}

\def\p{\pi}
\def\th{\theta}

\def\bar{\overline}
\def\l{\lambda}

\catcode`\@=11
  
\def\@normalsize{\@setsize\normalsize{15pt}\xiipt\@xiipt
\abovedisplayskip 14pt plus3pt minus3pt%

\belowdisplayskip \abovedisplayskip
\abovedisplayshortskip  \z@ plus3pt%
\belowdisplayshortskip  7pt plus3.5pt minus0pt}
\def\small{\@setsize\small{13.6pt}\xipt\@xipt
\abovedisplayskip 13pt plus3pt minus3pt%
\belowdisplayskip \abovedisplayskip
\abovedisplayshortskip  \z@ plus3pt%
\belowdisplayshortskip  7pt plus3.5pt minus0pt
\def\@listi{\parsep 4.5pt plus 2pt minus 1pt
            \itemsep \parsep
            \topsep 9pt plus 3pt minus 3pt}}
 
\def\underline#1{\relax\ifmmode\@@underline#1\else
        $\@@underline{\hbox{#1}}$\relax\fi}
\@twosidetrue
\relax

\catcode`@=12

\catcode`\@=11

\def\section{\@startsection{section}{1}{\z@}{3.5ex plus 1ex minus
   .2ex}{2.3ex plus .2ex}{\large\bf}}

\def\ps@headings{\def\@oddfoot{}\def\@evenfoot{}
\def\@oddhead{\hbox{}\hfill
        \makebox[.5\textwidth]{\raggedright\ignorespaces --\thepage{}--
        \hfill }}
\def\@evenhead{\@oddhead}
\def\subsectionmark##1{\markboth{##1}{}}
}
 
\ps@headings
 
\catcode`\@=12
 
\relax

\def\figcap{\section*{Figure Captions\markboth
        {FIGURECAPTIONS}{FIGURECAPTIONS}}\list
        {Fig. \arabic{enumi}:\hfill}{\settowidth\labelwidth{Fig. 999:}
        \leftmargin\labelwidth
        \advance\leftmargin\labelsep\usecounter{enumi}}}
 \relax
\def\tablecap{\section*{Table Captions\markboth
        {TABLECAPTIONS}{TABLECAPTIONS}}\list
        {Table \arabic{enumi}:\hfill}{\settowidth\labelwidth{Table 999:}
        \leftmargin\labelwidth
        \advance\leftmargin\labelsep\usecounter{enumi}}}
 \relax
\def\reflist{\section*{References\markboth
        {REFLIST}{REFLIST}}\list
        {[\arabic{enumi}]\hfill}{\settowidth\labelwidth{[999]}
        \leftmargin\labelwidth
        \advance\leftmargin\labelsep\usecounter{enumi}}}
 \relax
 
\catcode`\@=11
 
\def\marginnote#1{}
\newcount\hour
\newcount\minute
\newtoks\amorpm
\hour=\time\divide\hour by60
\minute=\time{\multiply\hour by60 \global\advance\minute by-
\hour}
\edef\standardtime{{\ifnum\hour<12 \global\amorpm={am}%
    \else\global\amorpm={pm}\advance\hour by-12 \fi
    \ifnum\hour=0 \hour=12 \fi
    \number\hour:\ifnum\minute<100\fi\number\minute\the\amorpm}}
\edef\militarytime{\number\hour:\ifnum\minute<100\fi\number\minute}
\def\draftlabel#1{{\@bsphack\if@filesw {\let\thepage\relax
  \xdef\@gtempa{\write\@auxout{\string
    \newlabel{#1}{{\@currentlabel}{\thepage}}}}}\@gtempa
    \if@nobreak \ifvmode\nobreak\fi\fi\fi\@esphack}
     \gdef\@eqnlabel{#1}}
\def\@eqnlabel{}
\def\@vacuum{}
\def\draftmarginnote#1{\marginpar{\raggedright\scriptsize\tt#1}}
\def\draft{\oddsidemargin -.5truein
        \def\@oddfoot{\sl preliminary draft \hfil
        \rm\thepage\hfil\sl\today\quad\militarytime}
        \let\@evenfoot\@oddfoot \overfullrule 3pt
        \let\label=\draftlabel
        \let\marginnote=\draftmarginnote
 
\def\@eqnnum{(\theequation)\rlap{\kern\marginparsep\tt\@eqnlabel}%
\global\let\@eqnlabel\@vacuum}  }
\def\preprint{\twocolumn\sloppy\flushbottom\parindent 1em
        \leftmargini 2em\leftmarginv .5em\leftmarginvi .5em
        \oddsidemargin -.5in    \evensidemargin -.5in
        \columnsep 15mm \footheight 0pt
        \textwidth 250mmin      \topmargin  -.4in
        \headheight 12pt \topskip .4in
        \textheight 175mm
        \footskip 0pt
 
\def\@oddhead{\thepage\hfil\addtocounter{page}{1}\thepage}
        \let\@evenhead\@oddhead \def\@oddfoot{} \def\@evenfoot{}
}
\def\titlepage{\@restonecolfalse\if@twocolumn\@restonecoltrue\onecolumn
     \else \newpage \fi \thispagestyle{empty}\c@page\z@
        \def\thefootnote{\fnsymbol{footnote}} }
\def\endtitlepage{\if@restonecol\twocolumn \else  \fi
        \def\thefootnote{\arabic{footnote}}
        \setcounter{footnote}{0}}  %\c@footnote\z@ }
\catcode`@=12
\relax
\def\ps@headings{\def\@oddfoot{}\def\@evenfoot{}
\def\@oddhead{\hbox{}\hfill
        \makebox[.5\textwidth]{\raggedright\ignorespaces --\thepage{}--
        \hfill }}
\def\@evenhead{\@oddhead}
\def\subsectionmark##1{\markboth{##1}{}}
}
 
\ps@headings
 
\relax

\newcommand{\newc}{\newcommand}
\newc{\ra}{\rightarrow}
\newc{\lra}{\leftrightarrow}
\newc{\beq}{\begin{equation}}
\newc{\eeq}{\end{equation}}
\newc{\bea}{\begin{eqnarray}}
\newc{\eea}{\end{eqnarray}}
\def\eps{\epsilon}
\def\la{\lambda}

\newc{\sm}{Standard Model}
\newc{\smd}{Standard Model}
\newc{\barr}{\begin{eqnarray}}
 \newc{\earr}{\end{eqnarray}}
  \newcommand{\ccaption}[2]{
    \begin{center}
    \parbox{0.85\textwidth}{
      \caption[#1]{\small{\it{#2}}}
      }
    \end{center}
    }

\ps@headings
 
\relax
 
\def\firstpage#1#2#3#4#5#6{
\begin{document}
%\draft
%\input epsf.tex

%%%%%%%%%%%
\begin{titlepage}
\nopagebreak
\title{\begin{flushright}
        \vspace*{-0.8in}
{ \normalsize  hep-ph/9808251 \\
ACT-7/98 \\
CERN-TH/98-216 \\
CTP-TAMU-25/98 \\
IOA-10/1998 \\
}
\end{flushright}
\vfill
{#3}}
\author{\large #4 \\[0.7cm] #5}
\maketitle
\vskip -7mm
\nopagebreak
\begin{abstract}
{\noindent #6}
\end{abstract}
\vfill
\begin{flushleft}
\rule{16.1cm}{0.2mm}\\[-3mm]

%%%%%%%%%%%
%November 1997
\end{flushleft}
\thispagestyle{empty}
\end{titlepage}}
 
\def\simlt{\stackrel{<}{{}_\sim}}
\def\simgt{\stackrel{>}{{}_\sim}}
\date{}
\firstpage{3118}{IC/95/34}
{\large\bf Neutrino Textures in the Light of Super-Kamiokande Data
and a Realistic String Model}
{J. Ellis$^{\,a}$,  G.K. Leontaris$^{\,a,b}$,
S. Lola$^{\,a}$ and D.V. Nanopoulos$^{\,c,d,e}$}%\\[-3mm]
{\normalsize\sl
$^a$Theory Division, CERN, CH 1211 Geneva 23, Switzerland\\[2.5mm]
\normalsize\sl
$^b$Theoretical Physics Division, Ioannina University,
GR-45110 Ioannina, Greece\\[2.5mm]
\normalsize\sl
$^c$Center for Theoretical Physics, Department of Physics,\\[-1.0mm]
\normalsize\sl
  Texas A\&M
 University, College Station, TX 77843 4242,  USA\\[2.5mm]
\normalsize\sl
$^d$Astroparticle Physics Group, Houston Advanced Research Center (HARC),
\\[-1.0mm]
\normalsize\sl
The Mitchell Campus, Woodlands, TX 77381, USA\\[2.5mm]
\normalsize\sl
$^e$ Academy of Athens, Chair of Theoretical Physics, Division of Natural
Sciences,\\[-1.0mm]
\normalsize\sl
 28 Panepistimiou Ave., Athens GR-10679,  Greece.}
{Motivated by the Super-Kamiokande
atmospheric neutrino data, we discuss possible textures for Majorana and
Dirac neutrino masses within the see-saw framework.
The main purposes of this paper are twofold: first to obtain
intuition from a purely phenomenological analysis, and secondly
to explore to what extent it may be realized in a specific model.
We comment initially on the simplified two-generation case,
emphasizing that large mixing is not incompatible with a large hierarchy
of mass eigenvalues.  We also emphasize that renormalization-group
effects
may amplify neutrino mixing, presenting semi-analytic expressions for
estimating this amplification. Several examples are then given of
three-family neutrino mass textures which may also accommodate the
persistent solar neutrino deficit, with different assumptions for the
neutrino Dirac mass matrices.
We comment on a few features of neutrino
mass textures arising
in models with a $U(1)$ flavour symmetry. Finally, we
discuss the possible pattern of neutrino masses in a `realistic' flipped
$SU(5)$ model derived from string theory, illustrating how a desirable
pattern of mixing may emerge. 
Both small- or large-angle MSW solutions are possible, whilst a
hierarchy
of neutrino masses appears more natural than near-degeneracy.
This model contains  some
unanticipated features
that may also be relevant in other models:
the neutrino Dirac matrices may not
be related closely to the quark mass matrices, and the heavy Majorana
states may include extra gauge-singlet fields.}

\setcounter{page}{0}

\section{Introduction}

There have recently been reports from the Super-Kamiokande
collaboration~\cite{SKam} and others~\cite{Kam}
indicating that the atmospheric neutrino deficit 
is due to neutrino oscillations. The data 
on electron events with visible energy
greater than 200 MeV are
in very good consistency with Standard Model expectations.
On the other hand, the number of events with muons 
is about half of the expected number, and the deficit becomes
more acute for larger values of $L/E$,
indicating that neutrino oscillations 
dilute the abundance of atmospheric $\nu_{\mu}$.
The possibility that $\nu_\mu \rightarrow
\nu_e$ oscillations
dominate is disfavoured by both Super-Kamiokande~\cite{SKam} and
CHOOZ data~\cite{chooz}. A fit to $\nu_{\mu}- \nu_{\tau}$ 
oscillations, with 
$\Delta m^2 = 5-50 \cdot 10^{-4}
 \; {\rm eV}^2$
and $\theta \sim \pi/4$ matches the data very well,
 but an admixture of
$\nu_\mu \rightarrow \nu_e$ oscillations cannot be excluded.

One intriguing feature of this scenario is the large mixing angle that
is required, and the question that arises is
how one could achieve this in
theoretically motivated models.
Large mixing angles in the neutrino
sector do arise naturally in a sub-class of GUT models
with flavour symmetries, as in~\cite{DG},
where they were used
to explain what was then only an ``atmospheric neutrino
anomaly''~\cite{Hir}.
Models with a single $U(1)$ symmetry
mostly predict small mixings~\cite{DLLRS},
principally because of the
constrained form of the Dirac mass matrices.
However, this need  not be a generic 
feature, as was shown in~\cite{LLR}. 
Many models 
with a single $U(1)$ symmetry
predict small mixings~\cite{DLLRS},
principally because of the
constrained form of the Dirac mass matrices.
However, this is also not a generic feature, and
textures with large $\nu_{\mu} -\nu_{\tau}$ 
mixing  have also been presented in~\cite{LLR}. Moreover,
string-derived models may well have 
a richer structure, with three or four
$U(1)$ symmetries.

However, models where the large
neutrino mixing arises from the Dirac mass
matrix may have a  problem with quark masses.
In many GUTs such as $SO(10)$, the neutrinos and up-type quarks
couple to the same Higgs and are in the
same multiplets, so their couplings arise from identical
GUT terms. Thus, in these cases
one would generate simultaneously
large mixing in the $u$-quark sector. 
Then, in order to obtain small mixing in $V_{CKM}$,
one needs to invoke some cancellation with mixing
in the $d$-quark sector. One way to overcome these difficulties
may be to invoke additional
symmetries, as arise in string-derived 
GUT models. In `realistic' models which also give the
correct pattern of quark masses and mixings, one can
hope to generate large neutrino mixing,
due to the combined form of the Dirac and
heavy Majorana mass matrices, even in cases where the
off-diagonal elements of the Dirac mass matrix are
not large by themselves. 
A study of phenomenologically viable 
heavy Majorana mass matrices leading to a large mixing angle,
for different  choices of the Dirac mass matrix,
has previously been presented in~\cite{ver}.~\footnote{Other textures with 
large mixing angles 
have also been proposed~\cite{recent,larmix,BLPR}.}

%As already mentioned, it is interesting to look for realistic neutrino
%mass
%textures
%in the framework of GUT models that can be motivated from
%string theory. 
Realistic string models have been in particular
constructed in the free-fermionic superstring formulation, with 
encouraging results.
Recently, due to better understanding
of non-perturbative string effects, which
may remove the previous apparent discrepancy between the string and gauge
unification scales, interest in string-motivated GUT
symmetries has been revived.
In this framework, we have looked recently~\cite{ELLN} at 
the predictions for quark  masses 
in the context of a flipped
$SU(5)\times U(1)$ model~\cite{aehn}, which is one of the 
three-generation superstring models derived in the free-fermion formulation.
The extension to lepton and neutrino
masses has various ambiguities,
since the original assignments~\cite{aehn} of the
lepton fields are not unique. Moreover, the model contains 
many singlet fields, and 
which of them develop non-zero vacuum expectation values (vev's) depends on the
choice of flat direction.

GUT and string models form the motivation for the analysis
contained in this paper. However, before addressing them, we first 
perform a more general phenomenological
analysis of neutrino masses and mixing, seeking to
understand the general message provided by the recent
data~\cite{SKam,Kam}. Equipped with this intuition, we then explore the
possibilities
for accommodating the data within specific models in which
the neutrino Dirac mass matrix is consistent with 
the charged-lepton and quark
mass matrices that we derived in~\cite{ELLN}. Some novel
features appear: flipped $SU(5)$ avoids the tight relation
between $u$-quark and neutrino Dirac mass matrices, and
gauge-singlet fields may be candidates for $\nu_R$ fields~\cite{GeoNan}.
Within this model, we prefer a hierarchy of neutrino masses, and may
obtain either the small- or the large-angle MSW solution to the
solar neutrino problem.

The layout of this paper is as follows. After a brief review in Section 2
of the data and their
implications, in Section 3 we analyze possible forms of the
Dirac and heavy Majorana mass matrices in a simplified
$2 \times 2$ model. Renormalization-group effects in this model
are studied in Section 4. Then, in Section 5 we explore certain
aspects of the multi-dimensional parameter space of
$3 \times 3$ models. Section 6 contains some comments on
models with $U(1)$ flavour symmetries. Finally, Section 7
studies neutrino mass matrices in the string model of~\cite{ELLN}
(which is reviewed in the Appendix), and Section 8 summarizes
our conclusions,
where we point to features that
may be generalizable to other models.

\section{Neutrino Data and their Implications}

The atmospheric neutrino data reported by Super-Kamiokande
and other experiments \cite{SKam,Kam} are explicable by

(a) $\nu_{\mu} \rightarrow \nu_{\tau}$ 
oscillations with
\begin{eqnarray}
\delta m^2_{\nu_\mu \nu_{\tau }}& \approx &
(10^{-2} \; {\rm to} \; 10^{-3})
\; {\rm eV^2}\\
sin^22\theta_{\mu \tau}&\ge  & 0.8
\end{eqnarray}
A description in terms of
$\nu_{\mu} \rightarrow \nu_{e} $
oscillations alone fits the data less well, and is in any
case largely excluded by the CHOOZ experiment \cite{chooz}.
However, there may be some admixture of $\nu_{\mu} - \nu_e$
oscillations (see, e.g.,
the last paper in \cite{recent}).

The solar neutrino data may be explicable in terms of
$\nu_e \rightarrow \nu_{\alpha}$ oscillations with either
\noindent
($b_1$) a small-angle MSW solution \cite{solar}
\begin{eqnarray}
\delta m^2_{\nu_e\nu_{\alpha}}&\approx &(3-10)\times 10^{-6}
\; {\rm eV^2}\\
sin^22\theta_{\alpha e} & \approx &  (0.4-1.3) \times 10^{-2}
\end{eqnarray}
or ($b_2$) a large-angle MSW solution 
\begin{eqnarray}
\delta m^2_{\nu_e\nu_{\alpha}}&\approx &(1-20)\times 10^{-5}
\; {\rm eV^2}\\
sin^22\theta_{\alpha e} & \approx &  (0.5-0.9)
\end{eqnarray}
or ($b_3$) vacuum oscillations 
\begin{eqnarray}
\delta m^2_{\nu_e\nu_{\alpha}}&\approx &(0.5-1.1)\times 10^{-10}
\; {\rm eV^2}\\
sin^22\theta_{\alpha e}&\ge  & 0.67
\end{eqnarray}
where $\alpha$ is $\mu$ or $\tau$.

One may also consider the possibility
(c) that there is a significant neutrino
contribution to the mass density of the Universe
in the form of hot dark matter, which would require
$\sum_i m_{\nu_i} \geq 3$ eV. If this was to be
the case, the atmospheric and solar neutrino data
would enforce
$m_{\nu_e} \approx m_{\nu_\mu} \approx m_{\nu_\tau} \geq 1$ eV.
This would be only marginally compatible with
$(\beta\beta)_{0\nu}$ limits,
which might require some
cancellations in the event of large mixing, as
required in scenarios $(b_2,b_3)$ above.
Motivation for a significant hot dark matter component
was provided some years ago by the need for some
epicycle in the standard cold dark matter model for
structure formation, in order to reconcile the
COBE data on fluctuations in the cosmic microwave
background radiation with
other astrophysical structure data~\cite{mixed}.
Alternative epicycles included a tilted spectrum
of primordial fluctuations and a cosmological
constant. In recent years, the case for mixed hot and cold
dark matter has not strengthened, whilst
recent data on large red-shift supernovae favour a
non-zero cosmological constant~\cite{Perlmutter}. 

Under these
circumstances, we consider abandoning the cosmological
requirement (c). In this case, the
atmospheric and solar neutrino conditions
(a,b) no longer impose near-degeneracy on
any pair of neutrinos, though this remains
a theoretical possibility.

Thus, one is led to consider the possibility of
a hierarchy of neutrino masses:
$m_{\nu_3} \gg m_{\nu_2},m_{\nu_1}$,
leaving open for the moment the
possibility of a second hierarchy
$m_{\nu_2} \gg m_{\nu_1}$. In either case,
condition (a) requires
$m_{\nu_3} \approx (10^{-1} \; {\rm to} \;
10^{-1 \frac{1}{2}})$ eV, and if
there is a second hierarchy
$m_{\nu_1} \ll m_{\nu_2} \approx (10^{-2} \; {\rm to} \;
10^{-3})$ eV~\footnote{We note that a sterile neutrino with
$\Delta m_{1,4}^2 \sim 1 eV^2$ is sometimes postulated in order
to accommodate the data from short-baseline neutrino
experiments. Such a possibility may be realized~\cite{ster}
within some variants
of the models we are examining, e.g., flipped SU(5), which include
additional light neutral singlets as well as
the three ordinary neutrinos. However, we do not discuss such
scenarios here.}. One may then wonder about the
magnitude of the mixing angles. It is well known
that large mixing is generic if off-diagonal entries
in the mass matrix are larger than differences between
diagonal entries. Can one reverse this
argument, i.e., to what extent is a large mixing angle incompatible
with a hierarchy of mass eigenstates
$m_{\nu_3} \gg m_{\nu_2}$? We study this question,
using, in the first place, a simple two-generation model.
Later we extend our analysis to the three-generation case,
and then examine whether the necessary mass
matrices have any chance of arising in models with
popular types of flavour symmetries, or in a model
derived from string theory.

\section{Mixing and Mass Hierarchies}

The light-neutrino mass matrix 
may be written as
\begin{equation}
m_{eff}=m^D_{\nu}\cdot (M_{\nu_R})^{-1}\cdot m^{D\dagger}_{\nu}
\label{eq:meff}
\end{equation}
where $m_{\nu}^D$ is the Dirac neutrino mass matrix and $M_{\nu_R}$
the heavy Majorana neutrino mass matrix. 
We consider initially generic forms for
$m_{\nu}^D$
and $M_{\nu_R}$, not forgetting that
many unified models with an $SO(10)$ structure give the relation
$m_{\nu}^D \sim m_u$. To identify which mass patterns
may fulfil the phenomenological
requirements outlined in the
previous section, we consider
an effective light-neutrino mass matrix 
with strong mixing. 
We then investigate which form of the heavy
Majorana mass matrix is compatible with a specific form 
of the neutrino Dirac mass matrix
\footnote{A classification of
the possible forms of the heavy Majorana mass matrices leading to
large mixing, for
various forms of the Dirac mass matrices,
was given previously in~\cite{ver}.}.

\subsection{Maximal Mixing and Hierarchical Masses in the 
 Two-Generation
Case}

For simplicity, we concentrate initially
on the $2 \times 2$ mass submatrix for the second and
third generations. According to (\ref{eq:meff}),
this may be written in the form 
\begin{equation}
M_{\nu_R} 
= m_{\nu}^{D\dagger}\cdot m_{eff}^{-1}\cdot m_{\nu}^D,
\label{mright}
\end{equation}
with
\begin{equation} 
m^{-1\ diag}_{eff} =
\left (
\begin{array}{ccc}
\frac{1}{m_{2}} & 0 \\
 0 & \frac{1}{m_{3}}
\end{array}
\right):
\;\;\;
m_{eff}^{-1} = V_{\nu} m_{eff}^{-1\ diag} V_{\nu}^T
\label{eq:done}
\end{equation}
where $V_{\nu}$ is the neutrino mixing matrix, and we are
going to explore large (23) mixing.
In most cases, there are small differences between the
mixing at the GUT scale and at low energies, so
we first focus on the possibility of obtaining the large mixing angle 
needed to resolve
the atmospheric neutrino problem directly
from the theory at high scales, discussing later possible
enhancement by renormalization-group effects at lower scales.
Parametrizing the $2 \times 2$ mixing matrix by

\beq
V_{\nu} = \left
(\begin{array}{ccc}
 c_{23} & -s_{23} \\
 s_{23} & c_{23}
\end{array}
\right)
\eeq
we see from (\ref{eq:done}) that $m^{-1}_{eff}$ has the form
\beq
m_{eff}^{-1} =\frac 1{m_2m_3} \left (
\begin{array}{cc}
{c^2_{23}}{m_3}+{s_{23}^2}{m_2} &
c_{23}s_{23}({m_3}-{m_2})
\\
c_{23}s_{23}({m_3}-{m_2})
& {c_{23}^2}{m_2}+{s^2_{23}}{m_3}
\end{array}
\right)
\;
\equiv
\;
d \left (
\begin{array}{cc}
 b/d & 1 \\
 1 & c/d
\end{array}
\right)
\label{eq:form}
\eeq
Identifying the entries
gives
\begin{equation}
\sin 2\theta_{23}  = 
2 d\frac{ m_2 m_3}{m_3 - m_2}, 
\label{angle}
\end{equation}
where the mass eigenvalues $m_{2,3}$ are given by
\begin{equation}
m_{2,3}  = 
\frac{2}{
b+c \pm 
\sqrt{(b-c)^2+4d^2}} 
\label{mevalues}
\end{equation}
and $\theta_{23}$ is the
$\nu_{\mu}-\nu_{\tau}$ mixing angle.
It is apparent from (\ref{mevalues}) that the two eigenmasses have the
same sign for
$bc > d^2$, whilst they have opposite signs if $bc< d^2$.

Substituting (\ref{mevalues}) into (\ref{angle}), we find that
\begin{equation}
\sin^2 2\theta_{23}  = 
\frac{4 d^2}{(b-c)^2 + 4 d^2}
\end{equation}
It is clear that maximal mixing: $\sin^2 2\theta_{23} \approx 1,\;
\theta_{23} \approx \pi/4$ is obtained 
whenever $|b-c| \ll |d|$. As seen in (\ref{mevalues}),
the ratio of the two mass
eigenvalues then depends on the ratio $(b + c)/d$,
and there is no particular reason to expect their
near-degeneracy, though this would occur if $|b,c| \ll |d|$.
%if $ b = c = 0$,
%leading to
%$m_3 = -m_2 = 1/d$.
%Of course, there must be at least some small
%splitting between the masses, but this could come entirely
%from the renormalization group runs, and thus
%will not concern us at this point.
%However, this is not the only
%solution with large mixing.
The fact that non-zero but similar values of $b$ and $c$ can
lead to large mixing with lifting of
the mass degeneracy
is indicated in Figs.~1 and 2.
In these figures, we plot
the mixing angle and the ratio of the eigenvalues
in terms of $c/d$, for $b/d = 0,0.25$ and $0.5$
respectively. The smaller values of $b/d$
correspond to the darker lines.

\begin{figure}
\vspace*{-3.1 cm}
\centerline{\epsfig{figure=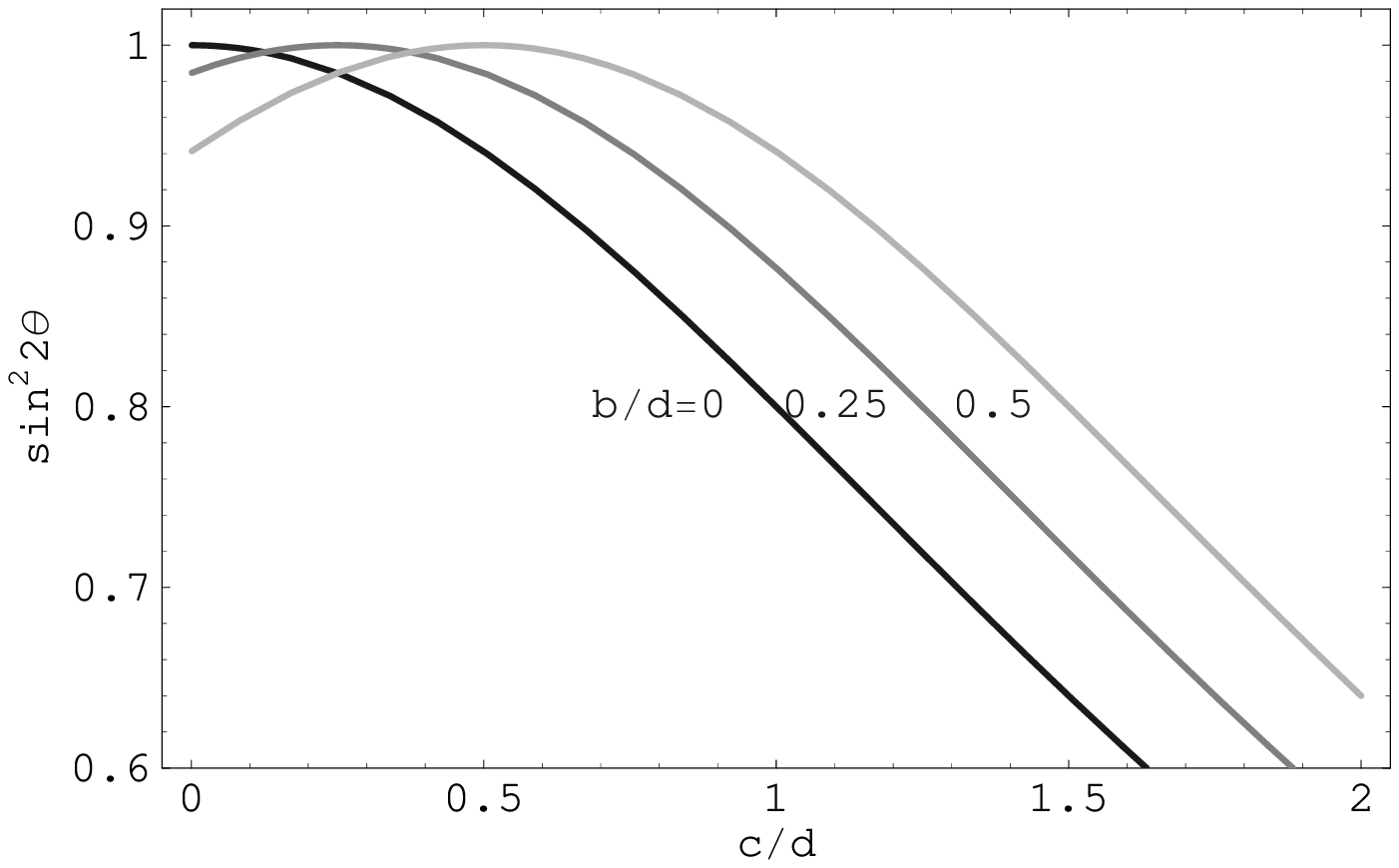,width=1.1\textwidth,clip=}}
\vspace*{-11.5 cm}
\ccaption{} {
The $2 \times 2$ mixing angle as a function
of the ratio $c/d$ in (\ref{eq:form}),
for selected values of $b/d$.
}
\end{figure}

\begin{figure}
\vspace*{-3.1 cm}
\centerline{\epsfig{figure=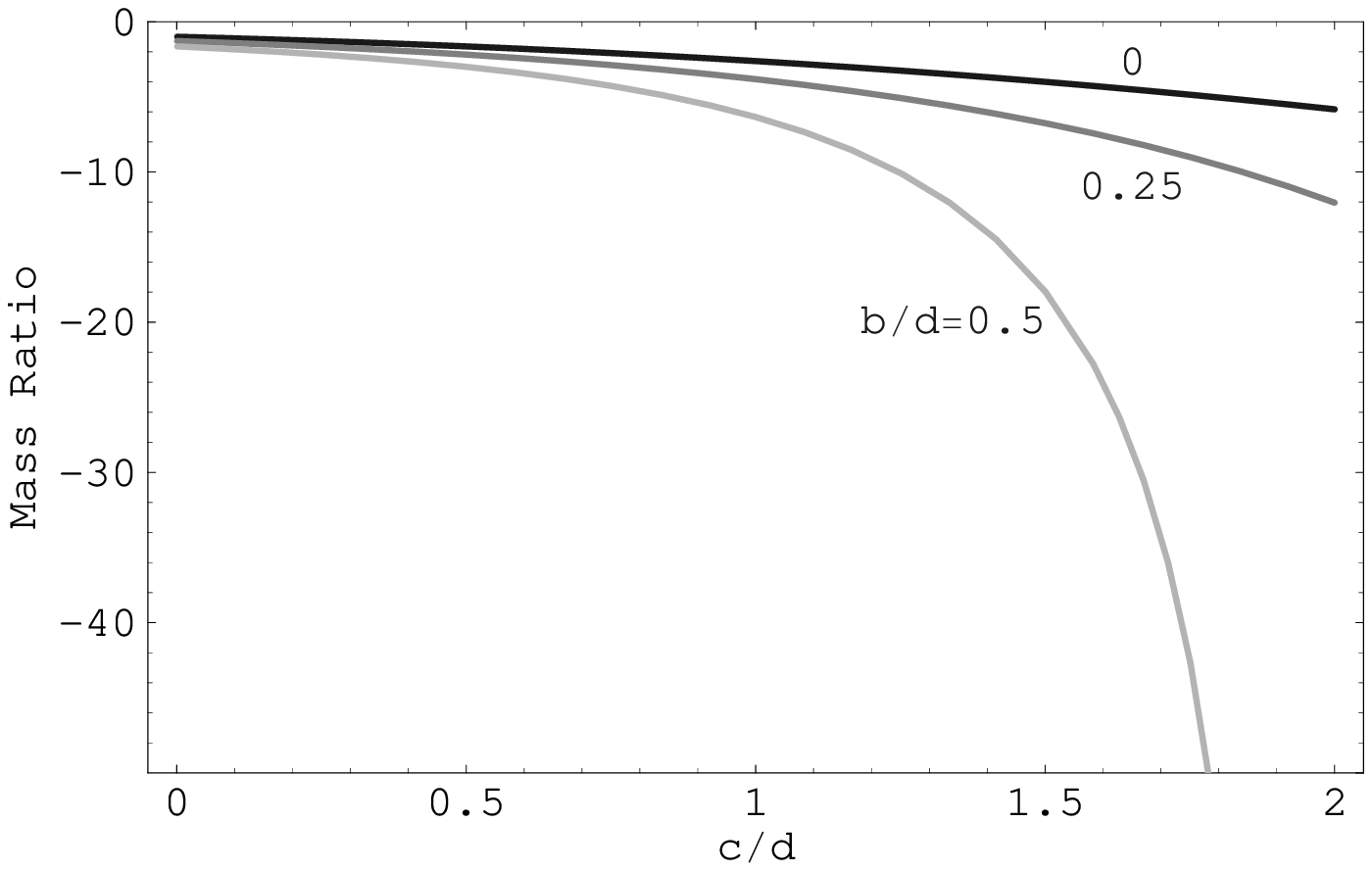,width=1.1\textwidth,clip=}}
\vspace*{-11.5 cm}
\ccaption{} {
The mass ratio $m_3/m_2$ as a function of
$c/d$ in (\ref{eq:form}),
for selected values of
$b/d$.
}
\end{figure}

As we see from these figures, 
the mixing angle can be sufficiently large: 
$\sin^2 2\theta_{23} \geq 0.8$
for a significant range of values
for $b$ and $c$. In particular, if the diagonal
entries are of the same order of
magnitude as the off-diagonal ones,
large mixing, which may even be amplified by
the renormalisation effects discussed below, is generated. 
Moreover, we observe that such large mixing 
does not require near-degenerate neutrinos,
but is also compatible with larger neutrino mass hierarchies.
If either $b$ or $c$ remain close to zero,
then the masses tend to be of comparable magnitude.
However, if both coefficients start becoming
as large as the off-diagonal entries, the
mass eigenvalues may spread out,
while the mixing angle may remain large.
Cancellations can arise automatically in the
calculation of the lighter mass eigenvalue,
as we subtract entries of comparable
magnitudes.
To illustrate this, let us look at a specific
set of values:
for $b=0.5,\; c=1.5$ and $d=1$,
sin$^2 2\theta_{23} = 0.8$.
However, the eigenvalues
are 2.1 and -0.1, differing by a factor of 20.

We conclude that the hierarchy $m_3 \approx 10^{-1}$ to $10^{-1.5}$ eV
$>> m_2 \approx 10^{-2.5}$ eV $>> m_1$ is compatible with a large
mixing angle $\sin^2 2\theta_{23}
 \ge 0.8$ as suggested by the
atmospheric and solar neutrino data. On the other hand,
obtaining the 1\% or better degeneracy between $m_2$ and $m_3$
that would be required in option (c) above, where there is
significant hot dark matter, would require a tighter
adjustment of parameters that appears less natural, if there is no
corresponding symmetry.

\subsection{Possible Textures of Dirac and Majorana Mass Matrices}

Having commented on the possible structure
of $m_{eff}$, the
next question is: from what forms of Dirac and heavy
Majorana mass structures may we obtain the
desired $m_{eff}$?
The form of the heavy Majorana mass
matrix $M_{\nu_R}$ 
may easily be found from (\ref{mright}),
once the neutrino Dirac mass matrix has been
specified. It is clear that if the 
neutrino Dirac mass matrix is diagonal,
one particular solution is 
\beq
M_{\nu_R} \propto (M_{\nu_R} )^{-1} \propto m_{eff} \propto
\left (
\begin{array}{cc}
0 & 1 \\
1 & 0 
\end{array}
\right )
\label{bel}
\eeq
Of course, as the Dirac mass matrix changes,
different forms of $M_{\nu_R}$ are required in order
to obtain the required form of $m_{eff}$.
This is exemplified in
Table 1, where we show the
textures that lead to 
$m_{eff}$ as given in (\ref{bel})
for various forms of symmetric Dirac mass matrices.
In the case of asymmetric mass matrices, one
would have more freedom in the choice of the
expansion parameters~\footnote{This freedom may, however, be 
limited if the mass patterns arise from
$U(1)$ symmetries, as we discuss subsequently.}, as seen in Table 2.
There we repeat the analysis of Table 1 for
the extreme case that one off-diagonal entry
of the $2 \times 2$ mass matrix is set to zero.

\begin{table}
\begin{center}
\begin{tabular}{|c|c|c|c|} \hline
$m_{\nu}^{D}$ & $(m_\nu^D)^{Diag}$ & 
$M_{\nu_R}$  & $M_{\nu_R}^{Diag}$  
\\ \hline \hline 
$\left(
\begin{array}{cc}
\la & \la^2 \\
 \la^2 & 1
\end{array}
\right)$ &
$\left(
\begin{array}{cc}
\la & 0\\
0 & 1
\end{array}
\right)$  &
$ M_N \left(
\begin{array}{ccc}
2 \la^2  & 1 \\
1 & 2 \la 
\end{array}
\right)$ &
$M_N \left(
\begin{array}{cc}
-1 & 0\\
0 & 1
\end{array}
\right)$  
\\ 
\hline
$\left(
\begin{array}{cc}
\la^2 & \la^2 \\
 \la^2 & 1
\end{array}
\right)$ &
$ \left(
\begin{array}{cc}
\la^2 & 0\\
0 & 1
\end{array}
\right)$  &
$ M_N \left(
\begin{array}{ccc}
2 \la^2  & 1 \\
1 & 2 
\end{array}
\right)$ &
$ M_N \left(
\begin{array}{cc}
(1-\sqrt{2}) & 0\\
0 & (1+\sqrt{2})
\end{array}
\right)$  
\\ 
\hline
$\left(
\begin{array}{cc}
\la^3 & \la^2 \\
\la^2 & 1
\end{array}
\right)$ &
$\left(
\begin{array}{cc}
\la^3 & 0 \\
0 & 1
\end{array}
\right)$ &
$ M_N \left(
\begin{array}{ccc}
2\la^3  & \la \\
\la & 2 
\end{array}
\right)$ &
$ M_N \left(
\begin{array}{ccc}
-\la^2/2  & 0 \\
0 & 2 
\end{array}
\right)$
\\ 
\hline
$\left(
\begin{array}{cc}
\la & 1 \\
 1 & \la
\end{array}
\right)$ &
$\left(
\begin{array}{cc}
-1 & 0 \\
 0 & 1
\end{array}
\right)$ &
$M_N \left(
\begin{array}{ccc}
2 \la  & 1 \\
1 & 2 \la 
\end{array}
\right)$ &
$M_N \left(
\begin{array}{ccc}
-1  & 0 \\
0 & 1  
\end{array}
\right)$
\\ 
\hline
$\left(
\begin{array}{cc}
\la^2 & 1 \\
 1 & \la
\end{array}
\right)$ &
$\left(
\begin{array}{cc}
-1 & 0 \\
0 & 1
\end{array}
\right)$ &
$ M_N \left(
\begin{array}{cc}
2 \la^2  & 1 \\
1 & 2\la 
\end{array}
\right)$ &
$ M_N \left(
\begin{array}{cc}
-1  & 0 \\
0 & 1
\end{array}
\right)$
\\ 
\hline
\end{tabular}
\end{center}

\vspace*{0.2 cm}
{\small Table 1: {\it Approximate forms for some of the
basic structures of symmetric
textures, keeping the dominant contributions.}}

\vspace*{0.4 cm}

\begin{center}
\begin{tabular}{|c|c|c|c|} \hline
$m_{\nu}^{D}$ & $(m_\nu^D)^{Diag}$ & 
$M_{\nu_R}$  & $M_{\nu_R}^{Diag}$  
\\ \hline \hline 
$\left(
\begin{array}{cc}
\la & 0 \\
 \la^2 & 1
\end{array}
\right)$ &
$\left(
\begin{array}{cc}
\la & 0\\
0 & 1
\end{array}
\right)$  &
$ M_N \left(
\begin{array}{ccc}
2 \la^2  & 1 \\
1 & 0
\end{array}
\right)$ &
$M_N \left(
\begin{array}{cc}
-1 & 0\\
0 & 1
\end{array}
\right)$  
\\ 
\hline
$\left(
\begin{array}{cc}
\la^2 & 0 \\
 \la^2 & 1
\end{array}
\right)$ &
$ \left(
\begin{array}{cc}
\la^2 & 0\\
0 & 1
\end{array}
\right)$  &
$ M_N \left(
\begin{array}{ccc}
2 \la^2  & 1 \\
1 & 0 
\end{array}
\right)$ &
$ M_N \left(
\begin{array}{cc}
-1 & 0\\
0 & 1
\end{array}
\right)$  
\\ 
\hline
$\left(
\begin{array}{cc}
\la^3 & 0 \\
\la^2 & 1
\end{array}
\right)$ &
$\left(
\begin{array}{cc}
\la^3 & 0 \\
0 & 1
\end{array}
\right)$ &
$ M_N \left(
\begin{array}{ccc}
2\la^2  & 1 \\
1 & 0
\end{array}
\right)$ &
$ M_N \left(
\begin{array}{ccc}
-1  & 0 \\
0 & 1
\end{array}
\right)$
\\ 
\hline
$\left(
\begin{array}{cc}
\la & 0 \\
 1 & \la
\end{array}
\right)$ &
$\left(
\begin{array}{cc}
\la^2 & 0 \\
 0 & 1
\end{array}
\right)$ &
$M_N \left(
\begin{array}{ccc}
2   & \la \\
\la & 0 
\end{array}
\right)$ &
$M_N \left(
\begin{array}{ccc}
2 & 0 \\
0 & -\la^2/2  
\end{array}
\right)$
\\ 
\hline
$\left(
\begin{array}{cc}
\la^2 & 0 \\
 1 & \la
\end{array}
\right)$ &
$\left(
\begin{array}{cc}
\la^3 & 0 \\
0 & 1
\end{array}
\right)$ &
$ M_N \left(
\begin{array}{cc}
2   & \la \\
\la & 0 
\end{array}
\right)$ &
$ M_N \left(
\begin{array}{cc}
2 & 0 \\
0 & -\la^2/2  
\end{array}
\right)$
\\ 
\hline
\end{tabular}
\end{center}

\vspace*{0.2 cm}

{\small Table 2: {\it Approximate forms for some of the
basic structures of asymmetric
textures, keeping the dominant contributions.}}
\label{table:maj}

\end{table}

Let us now make a few comments on
the tables, looking first at the case of symmetric Dirac
mass matrices. For the first texture, the
Dirac mass matrix is almost diagonal, so
a large mixing in the heavy Majorana sector
is directly communicated to $m_{eff}$.
In the third example, however, we see that
a large mixing angle in
the heavy Majorana sector may not lead 
to a large mixing in 
$m_{eff}$. In this case,
in order to obtain
a large mixing in $m_{eff}$,
we require a totally different
heavy Majorana mass texture, with the
larger element in the diagonal.
Let us now look at the third example
of the second table.
This texture is similar to the one we
just discussed,
with the exception of a zero in the (1,2) position.
The appearance of this zero brings us
back to the case where the large mixing in
the heavy Majorana sector is communicated
to $m_{eff}$. 
These observations, although
simple,  are of interest when we come to 
consider specific examples in the framework
of flavour symmetries in realistic models
at a later point in our discussion.

\subsection{Mixing-Angle Relations}

Equipped with these illustrative examples, we
now discuss in a more general way how
the mixing angles and mass hierarchies in the
various sectors are related, in particular by relaxing the
specific form (\ref{bel}) of $M_{\nu_R}$.
We consider the case of a symmetric Dirac mass matrix
with mixing angle $\vartheta$, define
$\phi$ to be the mixing angle in the heavy Majorana neutrino mass matrix,
and denote by $\theta$ the resulting mixing angle in 
the light-neutrino mass matrix $m_{eff}$~\footnote{We
drop for now the subindices
referring to the (2,3) sector of the neutrino matrices.}.
The heavy Majorana mass matrix can be
parametrised as
\bea
M_{\nu_R} = \left (
\begin{array}{cc}
M_2\cos^2\phi + M_3 \sin^2\phi &
(M_2-M_3)\cos\phi \sin\phi \\
(M_2-M_3)\cos\phi \sin\phi &
M_3\cos^2\phi + M_2 \sin^2\phi 
\end{array}
\right )
\eea
where the mixing angle 
is given by
\bea
\tan 2 \phi = 
\frac{ 
\sin (4\vartheta-2\theta)+r^2\sin 2\theta - 2 r R \sin 2 \vartheta
}
{
\cos (4\vartheta-2\theta)+r^2 \cos 2\theta - 2 r R \cos2 \vartheta
} 
\label{EQMIX}
\eea
Here, $M_3$ and $M_2$ are the eigenvalues of the
heavy Majorana mass matrix~\footnote{It is interesting to note that
(\ref{EQMIX})
exhibits a duality between $m_{eff}$ and 
$M_{\nu_R}$. Indeed, if one inverts the equation
$M_{\nu_R} 
= m_{\nu}^{D\dagger}\cdot m_{eff}^{-1}\cdot m_{\nu}^D$
to
$m_{eff}=m^D_{\nu}\cdot (M_{\nu_R})^{-1}\cdot m^{D\dagger}_{\nu}$,
one sees that $R$ and $\theta$ would stand for the relevant parameters
of $M_{\nu_R}$, while
$\phi$ would be the mixing angle in $m_{eff}$.},
$R \equiv
(m_2+m_3)/(m_3-m_2)$ where
$m_i$ are the eigenvalues of the light-neutrino mass matrix, and
$r \equiv
(m^D_2+m^D_3)/(m^D_3-m^D_2)$, where the $m^D_i$ are the
eigenvalues of the Dirac mass matrix.
In the limit where $m_2 = m_3$ (note that the signs have to be
the same) we have
$\tan 2 \phi = \tan 2 \vartheta $, while in the limit
$m_2^D = m_3^D$,
$\tan 2 \phi = \tan 2 \theta $.
Motivated by the equivalence at the unification scale of the $u$-quark 
and neutrino Dirac mass matrices in several GUTs,
in Fig.~3 we plot
$\sin^2 2\phi$ as a function of 
$\vartheta$ for 
$\theta = \pi/4$ and
$r = (180+1.4)/(180-1.4) \approx (m_t + m_c) / (m_t - m_c) \approx 1.01$,
for
three values of
$R$: 0, 1 and 10.
Fig.~4 shows the same plots, but for
$r = 3$.

\begin{figure}
\vspace*{-3.1 cm}
\centerline{\epsfig{figure=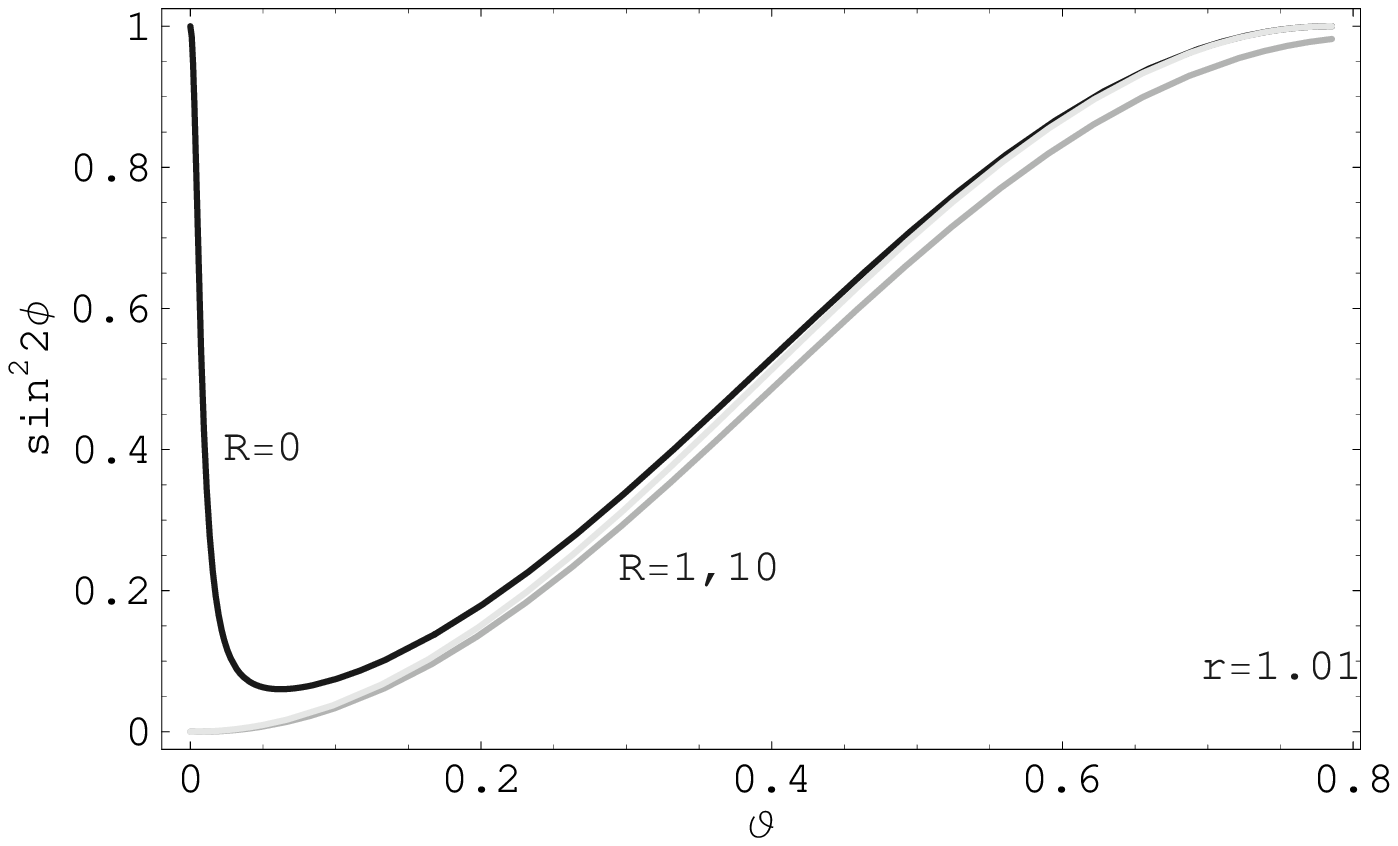,width=1.1\textwidth,clip=}}
\vspace*{-11.5 cm}
\ccaption{} {
The $2 \times 2$ heavy Majorana mixing angle as a function of the
mixing angle $\vartheta$ in $m_{\nu}^D$,
assuming a large hierarchy in its eigenvalues
$r \equiv (m^D_2+m^D_3)/(m^D_3-m^D_2)$,
for selected values of
$R \equiv
(m_2+m_3)/(m_3-m_2)$.
}
\end{figure}

\begin{figure}
\vspace*{-3.1 cm}
\centerline{\epsfig{figure=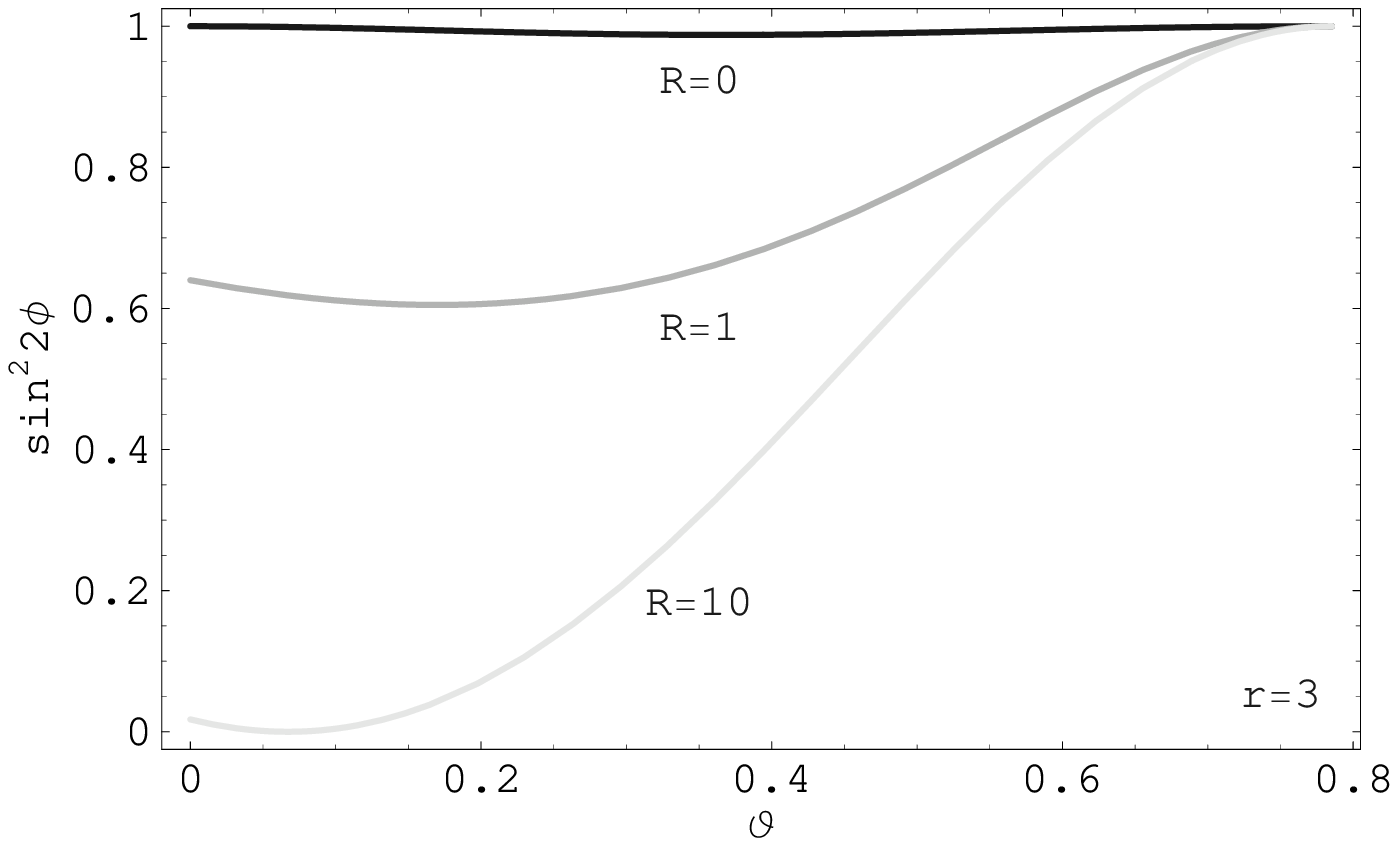,width=1.1\textwidth,clip=}}
\vspace*{-11.5 cm}
\ccaption{} {
As for Fig.3, assuming instead a small
hierarchy of eigenvalues in $m_{\nu}^D$.
}
\end{figure}

The parameters chosen for the plots
are representative of examples
with small and large hierarchies in
$m_{eff}$ and $m_{\nu}^D$: the choice
$R=0$ describes  the case with
$m_{2} = -m_{3}$, which we presented
in Tables 1 and 2. This structure arises
when the off-diagonal entries of
$m_{eff}$ are equal in magnitude, whilst being
much larger than the diagonal elements. The choice
$R =1$ represents examples with large hierarchies
(where $m_2$ can be neglected),
while $R=10$ describes a case with
$m_{2} = (9/11) m_{3}$, which is a typical example where
$m_2$ and $m_3$ are close in magnitude but
have the same sign. Concerning the Dirac neutrino
masses: the case with 
$r = (180+1.4)/(180-1.4) \approx 1.01$ is
typical of what one may expect 
in many unified (or partially unified) models,
where $m_{\nu}^D \sim m_{u}^D$. The choice
$r = 3$, on the other hand,
corresponds to the case where
$m_2^D = m_3^D/2$. This is a complementary example,
with small hierarchies in 
$m_{\nu}^D$.

In Fig.~3, we study a
case with a large hierarchy
in the Dirac mass matrix, much as could be
expected in
a wide number of unified or partially
unified models.
In this example, $R=0$ corresponds
to the case where the
two eigenvalues of $m_{eff}$ are
equal, but with opposite signs. 
This describes examples that appear in
Table 1.
In such a case,
maximal mixing in $M_{\nu_R}$ 
($\sin^2 2 \phi =1)$, is obtained for
a diagonal Dirac mass matrix
($\vartheta = 0$). 
{ This corresponds to the first
example of Table 1.}
For the same value of $R$, 
as $\vartheta$ starts increasing, a different form
of $M_{\nu_R}$ with a smaller (2,3) mixing
has a similar effect. This is 
what we see in example 3 of 
Table 1.

However, in the limit that
$\vartheta$ becomes quite large, 
we again reach a solution with a large mixing
in $M_{\nu_R}$. This is indicated in the examples 4 and 5 of
Table 1, where we see clearly this amplification
of the mixing in the heavy Majorana sector, 
as the off-diagonal entries of the
Dirac mass matrix become the dominant ones.
Notice also that example 2 indicates that when the
off-diagonal elements of the Dirac mass matrix
become of the order of the (1,1) element,
already the mixing in
$M_{\nu_R}$ has increased significantly.
On the other hand, we note that when the
Dirac neutrino mass matrix exhibits only a small hierarchy,
this picture is altered and $R=0$ leads generically to
a large mixing.
Indeed, again from Table 1, we see that
for $\lambda \approx O(1)$, the mixing in the
heavy Majorana mass matrix is always large.

In the limit when $R\ra 1$, $|m_2| \ll |m_3|$. Fig.~3 shows that,
in this case, maximal mixing in the light-neutrino sector 
can be obtained with negligible mixing in the heavy Majorana sector. 
Thus, for large hierarchies, even with a diagonal Dirac
matrix, we can get a large angle in $m_{eff}$
even for small mixing in the heavy
Majorana.  This result might at first seem rather surprising, 
but is related to the fact that the
solutions to the light-eigenvalue problem are quite sensitive.
It is also worth remembering that,
if $m_{eff}^{11} \approx m_{eff}^{22}$,
then the mixing angle is large, even if the
off-diagonal entries of $m_{eff}$ are very small.
Let us work out formulae (18,19) in this limiting case.
We denote by $m_{2,3}$ the light neutrino eigenmasses and 
assume maximal
mixing. 
In this case, the form of $m_{eff}$ is given by
\bea
m_{eff} 
 = \left (
\begin{array}{cc}
m_2 + m_3  &
m_2-m_3  \\
m_2-m_3 &
m_3 + m_2 
\end{array}
\right )
\eea  
and its inverse by
\bea
m_{eff} ^{-1}
 = 
\frac{1}{4 m_2 m_3}\left (
\begin{array}{cc}
m_2 + m_3  &
m_3 - m_2  \\
m_3-m_2 &
m_2 + m_3 
\end{array}
\right )
\eea  
If $m_c,m_t$ (equal to the quark masses at the unification scale)
are the entries in
the diagonal Dirac mass matrix, then the heavy Majorana mass matrix
that leads to maximal mixing is given by
\bea
M_{\nu_R} 
 \sim \left (
\begin{array}{cc}
(m_2 + m_3) m_c^2 &
(m_3-m_2) m_c m_t \\
(m_3-m_2)m_c m_t &
(m_3 + m_2) m_t^2 
\end{array}
\right )
\eea  
with a mixing  $\tan 2\phi = 2 (m_2+m_3) m_c^2/((m_3-m_2)(m_t^2-m_c^2))$.
For $m_3 \gg
m_2$ this mixing is indeed small
{ if the Dirac mass hierarchies 
are large}: $\tan 2\phi\sim m_c/m_t$.
{ However, if the Dirac mass hierarchies
are smaller, then the mixing angle 
increases in this case as well.
This we can see in more detail
in Fig.~4.}

Let us now comment on the sensitivity of this solution.
Suppose we keep $M_{\nu_R}$ as above, but modify
the second eigenvalue of the Dirac mass matrix to
$k m_t$. In this case, the mixing angle of $m_{eff}$
is found to be 
\bea
\sin^2 2\theta = \frac{4 k^2}{(1+k^2)^2}
\eea
For $k=1$,
we find sin$^2 2\theta = 1$, which falls to sin$^2 2\theta = 0.8$
for $k \approx 0.6$. 
Continuing to $k = 1/2$, we find
sin$^2 2\theta = 0.64$,
and for $k = 1/4$ we obtain
sin$^2 2\theta = 0.22$. We conclude that sin$^2 2\theta$ is in the
preferred region $\geq 0.8$ for quite a generic range of values of $k$.

To conclude this Section, we note that the case $R=10$ is also shown in
the Figs.~3, 4. We see that it
is similar to the previous one, but with smaller
sin$^2 2\phi$ than in the other cases, in the case of a small Dirac
mass hierarchy.

This phenomenological analysis indicates that solutions to the
atmospheric neutrino problem correlate 
and severely constrain
the masses and mixing hierarchies in the Dirac and heavy Majorana
sectors. We note that large mixing is not necessarily incompatible with 
a large hierarchy between two neutrino masses. The outcome of this
discussion can serve as a guideline in constructing realistic 
models of neutrino masses, as we do in section 7. 
Before that, however, we
analyze possible modifications of the above results due to
renormalization-group effects, and then we discuss how this $2 \times 2$
analysis may be embedded in a fuller $3 \times 3$ analysis.

%\begin{figure}
%\vspace*{-2 cm}
%\epsfxsize=18cm \epsfbox{Fig2.eps}
%\end{figure}

\section{Renormalization-Group Effects}

Up to now, we have
discussed the situation in which a maximal (23) mixing
angle appears already at the GUT scale. However,
this is not the only possibility.
Within the minimal supersymmetric extension of the Standard Model
(MSSM), it has been found that
renormalisation group effects may
amplify the mixing~\cite{ang1,ang2}. 
Whether this happens depends on the magnitude of
$h_{\tau}$: for large
$h_{\tau}$, for which large large tan$\beta$ is necessary, the
(23) and (13) mixing angles may be amplified significantly.
For some examples, initial  values of
sin$\theta_{23} \geq 0.3$
can lead to large mixing at
low energies~\cite{tanim,nurges}, and similar results
have been found for the (13) mixing at 
large tan$\beta$. On the other hand, the (12) mixing remains
essentially unchanged even for large tan$\beta$, and
renormalisation-group effects
can be neglected for small tan$\beta$.
We now discuss such renormalization-group effects in more detail.

\subsection{Renormalization-Group Equations}

Between the GUT scale and the scale of the
heavy Majorana neutrinos, $M_{N}$,
there is an effect on the mixing angle
due to the running~\footnote{We work in this
paper at the one-loop level.} of the
Dirac neutrino coupling $Y_N$:
 \begin{eqnarray}
  8\p^2 {d\over dt}({ Y}_N { Y}_N^\dagger)& = &
 \{ -\sum_i c_N^i g_i^2+3 ({ Y}_N { Y}_N^\dagger)
+{\rm Tr}[3({ Y}_U { Y}_U^\dagger)+({ Y}_N { Y}_N^\dagger)] \}
({ Y}_N { Y}_N^\dagger)  \nonumber \\
       & &\qquad \qquad   +{1\over 2}\{({ Y}_E { Y}_E^\dagger)
 ({ Y}_N { Y}_N^\dagger)+({ Y}_N { Y}_N^\dagger)
({ Y}_E { Y}_E^\dagger)\}  
 \end{eqnarray}
where
$t$ is the logarithmic renormalization-group scale,
$c^i_N = (3/5, 3, 0)$ for the MSSM,
and we denote the Dirac couplings of other types of fermion $F$ by $Y_F$.
We see from these equations that
the various entries of $m^{D}$ 
run differently:
large Yukawa couplings, which lower $Y_N$,
have a bigger effect on 
$m^{D}_{33}$  than on the rest of the 
elements. This alters
the structure of the Dirac mass matrix,
in turn affecting the magnitude
of the mixing angle. The effect  
becomes more relevant in examples where
cancellations between various entries 
may lead to amplified mixing in $m_{eff}$.

Below the right-handed Majorana mass scale, 
${ Y}_N$ decouples and the relevant running
is that of the effective neutrino mass operator~\cite{ang2}:
 \begin{equation} 
    8\p^2 {d\over dt}{ \k} = \{-({3\over 5} g_1^2+3 g_2^2)+
{\rm Tr} [3{ Y}_U { Y}_U^\dagger]\}\k
 +{1\over 2}\{({ Y}_E { Y}_E^\dagger)\k +
   \k ({ Y}_E { Y}_E^\dagger)^T\}       \ ,
\label{MASSES}
\end{equation}
which we use later to study the variation in the diagonal
entries in $m_{eff}$. Off-diagonal entries enter into
the neutrino mixing angle $\th_{23}$, whose running
is given by \cite{ang2}:
 \begin{equation}
    16\p^2 {d\over dt}{\sin^2 2\th_{23}} =-2\sin^2 2\th_{23}
        (1-\sin^2 2\th_{23})
 (Y_{E3}^2-Y_{E2}^2){\k^{33}+\k^{22}\over
\k^{33}-\k^{22}}  \ 
\label{MIXING}
 \end{equation}
As this equation indicates, sin$^2 2\th_{23}$ 
may be particularly strongly affected as one runs down
from the GUT to the electroweak scale
(i) if $Y_{E3}$ is large, and (ii)
if the diagonal entries of $m_{eff}$ are 
close in magnitude.
Thus the exact 
evolution of the
mixing angle depends on the
particular texture being studied.

In general, the amplification effects
that one may obtain for large
tan$\beta$ are of the order of
30 - 50\%, for cases where one starts
with a small or moderate value of the
mixing angle at the GUT scale.
However, in particular combinations
of textures for the Dirac and the heavy Majorana mass
matrices, cancellations between various terms may
lead to even larger amplifications of the mixing angles.
It has been noted in such cases that the
running of the Yukawa couplings between the
GUT scale and the effective $M_N$ scale
may strengthen such cancellation
effects, thus increasing the 
mixing significantly \cite{tanim}.
A similar effect may arise below $M_N$.
Examination of the equation (\ref{MIXING}) that describes the 
running of the mixing angle indicates that
significant amplification may be 
obtained for textures where
$m_{eff}^{22}$ and $m_{eff}^{33}$
are close in magnitude \cite{ang2}.

\subsection{Semi-Analytic Solutions}

In order to get a better feeling for the
magnitude of the mixing angle,
it is useful to look for semi-analytic solutions to 
one-loop equations (\ref{MASSES},\ref{MIXING}).
To do so, we start with the differential
equations for the diagonal elements
of  the effective neutrino mass matrix. These are given by
\bea
\frac{1}{m_{eff}^{22}}\frac{d}{d t}m_{eff}^{22}&=& 
\frac{1}{8\pi^2}\left( -c_i g_i^2 + 3 h_t^2 \right)
 \\
\frac{1}{m_{eff}^{33}}\frac{d}{d t}m_{eff}^{33}&=& 
\frac{1}{8\pi^2}\left( -c_i g_i^2 + 3 h_t^2 + h_{\tau}^2\right)
\eea
For the $m_{eff}^{33}$ element, simple integration yields
\bea
\frac{m_{eff}^{33}}{m_{eff,0}^{33}}&=& 
 exp\left\{ \frac{1}{8\pi^2}\int_{t_0}^t
             \left(-c_i g_i^2 + 3 h_t^2 +h_{\tau}^2\right)\right\}
\nonumber\\
    &=& I_g\cdot I_t \cdot I_{\tau}
\eea
where
\bea
 I_g& =& exp[\frac{1}{8\pi^2}\int_{t_0}^t(-c_i g_i^2 dt)]\\
I_t &=& exp[\frac{1}{8\pi^2}\int_{t_0}^t  h_t^2 dt]\\
    I_{\tau} &=& exp[\frac 1{8\pi^2}\int_{t_0}^t h_{\tau}^2 dt]
\eea
and $ m_{eff,0}^{33}$ is the initial
condition.
This condition is defined at
$M_N$, at the stage when $h_N$ decouples from
the renormalisation-group equations. For
simplicity of presentation, we assume for the sake of the
following discussion that $M_N \approx M_{GUT}$.
Similarly, we find that:
\bea
    m_{eff}^{22}& =& m_{eff,0}^{22}\cdot I_g\cdot I_t \\
\eea
so that
 $m_{eff}^{33}/m_{eff}^{22} = I_{\tau}\cdot
m_{eff,0}^{33}/m_{eff,0}^{22}$,
leading to the formula
\bea
   \frac{m_{eff}^{33}+m_{eff}^{22}}{m_{eff}^{33}-m_{eff}^{22}}& =&
    \frac{m_{eff,0}^{33} I_{\tau} +m_{eff,0}^{22}}
{m_{eff,0}^{33} I_{\tau}-m_{eff,0}^{22}} \equiv  f(I_{\tau})
\eea
for the diagonal mass-matrix elements.

We can then convert
the one-loop evolution equation (\ref{MIXING}) for sin$^2\theta$ to a 
differential equation for $T = \tan^22\theta$:
\bea
\frac{d T}{T} &=& 
-\frac{2}{16\pi^2} h_{\tau}^2 f(I_{\tau})
\label{tanev}
\eea
The solution to (\ref{tanev}) is
\bea
    \tan^2 2\theta &=& \tan^2 2\theta_0 I_2(h_{\tau})
\label{tan}
\eea
with 
\bea
I_2(h_{\tau}) =  exp\left\{ -\frac{1}{8\pi^2} \int_{t_0}^t
                  h_{\tau}^2 f(I_{\tau})\right\} 
\eea
We see that the only parameters which enter into the final
formula are the initial conditions and 
an integral that incorporates
all the renormalization-group running
of $h_{\tau}$.

\subsection{Some Implications}

The following are the most important deductions we
extract from the above equations.
Suppose we start with a generic 
$m_{eff}$ at a high scale: then,
$m_{eff}^{33}$ decreases
more rapidly than
$m_{eff}^{22}$, due to the effect of the
$\tau$ Yukawa coupling. If one starts with
$m_{eff}^{22}$ and $m_{eff}^{33}$ 
relatively close in magnitude,
the expectation is that at a given scale
they may become equal, in which case the mixing angle is maximal.
How fast this happens, depends on the
magnitude of $h_{\tau}$. The larger
$h_{\tau}$,  the earlier the entries
may become equal. Of course, $h_{\tau}$ also
decreases while running down to low energies, 
and this has also to be taken into account.
The scale where the mixing angle is maximal
is given by the relation
\bea
I_{\tau} = \frac{m_{eff,0}^{22}}{m_{eff,0}^{33}}
\eea
After reaching the maximal angle at some
intermediate scale, the running of
$h_{\tau}$ results in
\bea
{m_{eff,0}^{33}}  < {m_{eff,0}^{22}} 
\nonumber
\eea
This changes the sign of
$f(I_{\tau})$ and results in a rather rapid decrease of
the mixing. 
In order, therefore, for a texture of this type to be
viable, there needs to be a balance between
the magnitudes of $h_{\tau}$ and
$m_{eff}^{33}-m_{eff}^{22}$ at the GUT
scale.  If the splitting is small and
the coupling large, then the maximal value 
for the mixing will be obtained too early
to survive at low energies. 

Let us explore the circumstances under which the mixing
becomes close to maximal at low scales.
We consider an example where
$h_{\tau} = h_{t} 
= h_b \equiv h = 3$,
$M_{GUT} = 1.1 \times 10^{16}$ GeV
and the common gauge coupling at the unification scale
is $0.039$. Also, we take
the scale of supersymmetry breaking to be
around 1 TeV. We find that a texture~\cite{ang2}
with
\bea
m_{eff}^{22} = 0.6, \;\;\;
m_{eff}^{23} = 0.035, \;\;\;
m_{eff}^{33} = 1.0,
\eea
which has a starting value for the
mixing given by
sin$^2 2\theta \approx 0.03$,
reaches {\it maximal mixing}: sin$^2 2\theta \sim 1$ at
$\approx 1$ TeV.
If we assume the same texture but take
$h = 2$, the
mixing is only  
sin$^2 2\theta \approx 0.35$
at $\approx 1$ TeV. However, for the
same coupling, making the modification to
$m_{eff}^{22} = 0.7$
leads to maximal mixing at around 5.5 TeV.
Finally, for
$h = 1$ at the GUT scale, we need
to modify 
$m_{eff}^{22}$ to approximately 0.8,
in order for maximal mixing to occur around the
TeV scale.

Let us also look at another specific example texture.
We assume the texture 
\beq
m_{eff} = \left
(\begin{array}{cc}
1-x & x^2 \\
x^2 & 1+x 
\end{array}
\right)
\eeq
with $x \approx 0.2$, so that the off-diagonal elements
are much smaller than the mass splitting of
the diagonal ones. We are going to see
that renormalisation group effects
will lead to a very large increase
of the mixing angle.
In this case, the one-loop 
running of the mixing
angle is indicated in Fig.~5.
Here we took as initial conditions: 
$M_{N} \approx M_{GUT} = 10^{16}$ GeV,
a common coupling at the unification scale
$0.042$ and $h_{t} = h_{b} = h_{\tau} = 2.0$.
This running indicates that,
for this example, the mixing has indeed changed 
significantly as we run down to lower energies.

\begin{figure}[t]
\vspace*{-3.1 cm}
\centerline{\epsfig{figure=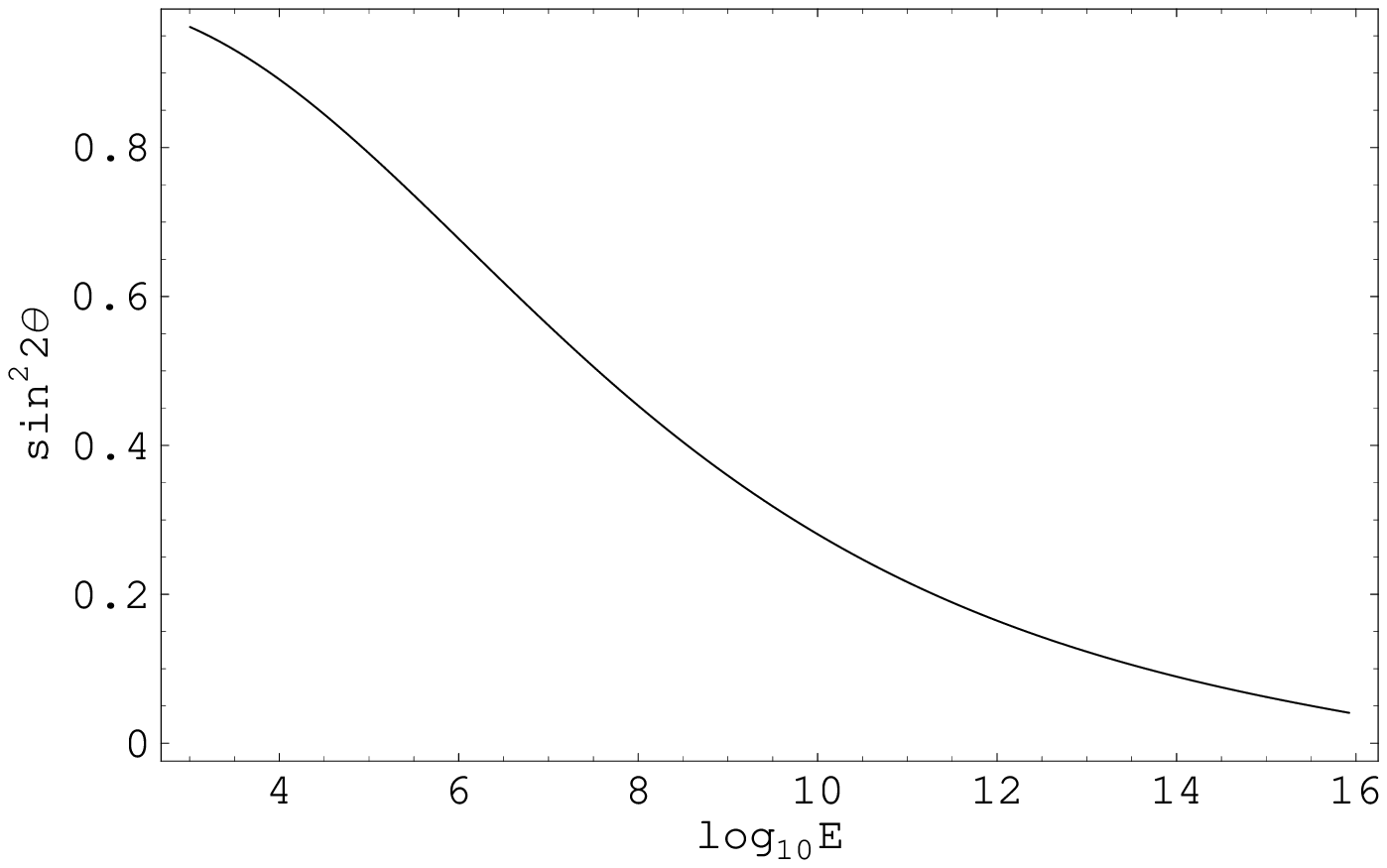,width=1.1\textwidth,clip=}}
\vspace*{-11.5 cm}
\ccaption{}{
An example of the renormalisation-group enhancement of the
$2 \times 2$ light-neutrino
mixing angle, starting from a small value at the GUT scale.
We assume initial Yukawa couplings
$h_{t} = h_{b} = h_{\tau} = 2.0$,
corresponding to a large value of $tan\beta$.
}
\end{figure}

%%%%%%%%%%%

\section{Sample Textures in Three-Generation Examples}

So far, we have worked in the limit
where the solar neutrino problem is
resolved by a small mixing angle.
However, this need not be the case, and one
should consider what happens if this mixing
is also large~\footnote{We also note that a hybrid solution 
involving both
resonance transitions and vacuum oscillations,
with intermediate values of the mixing angle,
has been proposed \cite{LP97},
and solutions consistent with realistic models
have been explored \cite{georg}.}.
In this case, we need to consider the general $3\times 3$ mixing
problem. We can clearly proceed as in the case
of 2 $\times 2$ mixing, and investigate
the relations between the
mixing angles and hierarchies in 
the Dirac, heavy and light Majorana mass matrices.
However, the number of parameters is very large,
and one cannot proceed far without 
making assumptions on the patterns of mixing and the structure of the mass
matrices. We write the generic form of a $3 \times 3$ mixing-angle matrix
(ignoring phases) in the form
\begin{equation}
V_{3 \times 3} = \left
(\begin{array}{ccc}
c_{12} & -s_{12} c_{13} & -s_{12} s_{13} \\
s_{12} c_{23} & c_{12}c_{23}c_{13}+s_{23}s_{13} & c_{12}c_{23}s_{13}-s_{23}c_{13} \\
s_{12}s_{23} & c_{12}s_{23}c_{13}-c_{23}s_{13} & c_{12}s_{23}s_{13} + c_{23}c_{13}
\end{array}
\right)
\end{equation}
where $s_{ij},c_{ij}$ stand for 
$\sin\theta_{ij}$ and $\cos\theta_{ij}$,
respectively, and will explore the implications of various
possible hierarchies between the angles $\theta_{ij}$.
Investigating the 
possible hierarchies
within $m_{eff}$ is then straightforward.

\subsection{Cases with Maximal Mixing}

We first assume for simplicity of discussion that
$\theta_{23}$ is maximal~\footnote{We saw, however, in the previous
Section
that for large $h_{\tau}$ the
(23) mixing angles (and similarly the (13) angle) can be significantly
modified.},
and that $\theta_{12} \gg \theta_{13}
\sim 0$. In this case,
$m_{eff} = V_{3 \times 3}.m_{eff}^{diag}. V_{3 \times 3}^{\dagger}$ is
given by
\beq
m_{eff} = \left
(\begin{array}{ccc}
m_1 c_{12}^2 + m_2 s_{12}^2 &
\frac{(m_1-m_2)c_{12}s_{12}}{\sqrt{2}} &
\frac{(m_1-m_2)c_{12}s_{12}}{\sqrt{2}} \\
\frac{(m_1-m_2)c_{12}s_{12}}{\sqrt{2}} &
\frac{1}{2} (m_3 + m_1 s_{12}^2 + m_2 c_{12}^2) &
\frac{1}{2} (-m_3 + m_1 s_{12}^2 + m_2 c_{12}^2)  \\
\frac{(m_1-m_2)c_{12}s_{12}}{\sqrt{2}} &
\frac{1}{2} (-m_3 + m_1 s_{12}^2 + m_2 c_{12}^2)  &
\frac{1}{2} (m_3 + m_1 s_{12}^2 + m_2 c_{12}^2)  
\end{array}
\right)
\eeq
and one may look at the implications for
mass hierarchies. Initially, we prefer to simplify further to the
case of maximal $\theta_{12},\theta_{23}$ mixing. In this case,
\beq
m_{eff} = \left
(\begin{array}{ccc}
\frac{m_1 + m_2}{2} & \frac{m_1 - m_2}{2\sqrt{2}} 
& \frac{m_1 - m_2}{2\sqrt{2}} \\
\frac{m_1 - m_2}{2\sqrt{2}} &\frac{1}{4} (m_1+m_2+2m_3) &
\frac{1}{4} (m_1+m_2-2m_3) \\
\frac{m_1 - m_2}{2\sqrt{2}} &
\frac{1}{4} (m_1+m_2-2m_3) & 
 \frac{1}{4} (m_1+m_2+2m_3) 
\end{array}
\right)
\label{maxmix}
\eeq
We saw in Section 3 that,
when all the entries of a 
$2 \times 2$ matrix are of the same
order of magnitude, plausible cancellations may still lead
to large hierarchies between the eigenvalues,
even in the presence of a large mixing.
We can visualize the type of texture of $3 \times 3$ $m_{eff}$ 
(\ref{maxmix}) that 
is consistent with such maximal mixing by considering
specific limiting cases for the $m_i$. 

For example,
in the limit $m_3 \gg m_2 \gg m_1$, one has
\beq
m_{eff} = {m_3 \over 2} \left(
\begin{array}{ccc}
0 & 0 & 0 \\
0 & 1 & -1 \\
0 & -1 & 1
\end{array}
\right) + {m_2 \over 2} \left(
\begin{array}{ccc}
1 & -{1 \over \sqrt{2}} & -{1 \over \sqrt{2}} \\
-{1 \over \sqrt{2}} & {1 \over 2} & {1 \over 2} \\
-{1 \over \sqrt{2}} & {1 \over 2} & {1 \over 2}
\end{array}
\right) + {m_1 \over 2} \left(
\begin{array}{ccc}
1 & {1 \over \sqrt{2}} & {1 \over \sqrt{2}} \\
{1 \over \sqrt{2}} & {1 \over 2} & {1 \over 2} \\
{1 \over \sqrt{2}} & {1 \over 2} & {1 \over 2}
\end{array}
\right)
\label{expansion}
\eeq
On the other hand, if one considers near-degenerate cases
$m \equiv m_3 \sim m_2 \sim m_1 : \Delta \equiv |m_3| - |m_2| \gg \delta 
\equiv |m_2| - |m_1|$, 
there are various possibilities, distinguished by
the relative signs of the eigenvalues. For example,
if all the eigenvalues have the same sign, 
one finds the texture:
\beq
m_{eff} = m_2 \left(
\begin{array}{ccc}
1 & 0 & 0 \\
0 & 1 & 0 \\
0 & 0 & 1
\end{array}
\right) + \frac{\Delta}{2} \left(
\begin{array}{ccc}
0 & 0 & 0 \\
0 & 1 & -1 \\
0 & -1 & 1 
\end{array}
\right) - \frac{\delta}{2} \left(
\begin{array}{ccc}
1 & {1 \over \sqrt{2}} & {1 \over \sqrt{2}} \\
{1 \over \sqrt{2}} & {1 \over 2} & {1 \over 2} \\
{1 \over \sqrt{2}} & {1 \over 2} & {1 \over 2} 
\end{array}
\right)
\eeq    
On the other hand, 
if one of the eigenvalues has a different sign from the
other two, this structures get modified. Suppose, for example, that
$m_1$ and $m_3$ are positive and
$m_2$ negative. We the find
\beq
m_{eff} = \frac{m_2}{2} \left(
\begin{array}{ccc}
0 &{-2 \over \sqrt{2}}&{-2 \over \sqrt{2}} \\
{-2 \over \sqrt{2}}  & -1 & 1 \\
{-2 \over \sqrt{2}} &  1 & -1
\end{array}
\right) + \frac{\Delta}{2} \left(
\begin{array}{ccc}
0 & 0 & 0 \\
0 & 1 & -1 \\
0 & -1 & 1 
\end{array}
\right) - \frac{\delta}{2} \left(
\begin{array}{ccc}
1 & {1 \over \sqrt{2}} & {1 \over \sqrt{2}} \\
{1 \over \sqrt{2}} & {1 \over 2} & {1 \over 2} \\
{1 \over \sqrt{2}} & {1 \over 2} & {1 \over 2} 
\end{array}
\right)
\eeq    

We return later to these suggestive examples, but 
first discuss how $m_{eff}$ may be derived
from the primary Dirac and Majorana mass matrices of the fundamental
theory, which may be some GUT and/or string model.

\subsection{Heavy Majorana Mass Textures with Matched Mixing}

Up to now, we have been discussing possible 
forms of $m_{eff}$ that
are consistent with the 
atmospheric and solar neutrino
data. 
However, in a more fundamental model, the gauge group
structure, as well as possible $U(1)$ symmetries and string
selection rules, predict the structure of the Dirac and 
heavy Majorana matrices, while 
$m_{eff}$ is a secondary output of the
see-saw mechanism.
Thus, to make contact with such unified (or partially unified) theories, 
it is essential to analyze the forms of Dirac and heavy
Majorana mass matrices that are suggested by the experimental data.
Such an analysis may reveal
relations
between the mass and mixing hierarchies
of the different neutrino sectors that
can then be used as a guideline in
investigations that involve realistic models, as we discuss in Section 7.

In general, calculating the heavy Majorana mass matrix
involves 12 parameters: the 6 eigenvalues
of $m_{\nu}^{D}$ and $m_{eff}$ and
3 mixing angles for each of these two
matrices. General formulae for all the entries in the
full $3 \times 3$ $M_{\nu_R}$ matrix in terms of all these parameters
are easily derived but 
quite complicated, and are not given here. Instead,
we look at some limiting cases. It is convenient to
parametrize these cases in terms of the hierarchy factors
$x \equiv m_1 / m_3, y \equiv m_2 / m_3$ for 
the ratios of eigenvalues of $m_{eff}$
and $\lambda_1 \equiv M_{\nu_R,1} / M_{\nu_R,3},
\lambda_2 \equiv M_{\nu_2}^D / M_{\nu_3}^D$ for
the ratios of eigenvalues of the neutrino Dirac mass
matrix $m_{\nu}^D$.

We initially
assume one large mixing angle in the effective light Majorana
matrix. Then, we can distinguish two cases for
the structure of the heavy Majorana matrix.
The first is that of {\it matched mixing}, when there is no large mixing
in other sectors of either
the light Majorana or the Dirac matrices, in which case the
problem is equivalent to the $2\times 2$ case considered previously.
In particular, consider
first the possibility that the Dirac mass matrix is 
diagonal to a good approximation.
Then, the form of the heavy Majorana mass 
matrix becomes
\beq
M_{\nu_R} \propto 
\left
(\begin{array}{ccc}
\l_1^2\over x & 0 & 0 \\ 0 & \l_2^2(y+1)\over 2 y & \l_2(y-1)\over  2 y \\
0 & \l_2(y-1)\over 2 y & (y+1)\over 2 y
\end{array}
\right)
\label{twobytwo}
\eeq
In the particular case that $y= m_2 / m_3 = -1$,
this leads to a texture of the form
\beq
M_{\nu_R} \propto  \left
(\begin{array}{ccc}
\l_1^2\over  x & 0 & 0 \\
0 & 0 & \l_2 \\
0 & \l_2 & 0
\end{array}
\right)
\label{yminusone}
\eeq
which resembles one of the popular $2 \times 2$
textures in Table 1.

Alternatively, for large hierarchies in $m_{eff}$,
i.e., for small mass ratios;
$x \ll y \ll 1$, the
form of the heavy Majorana mass 
matrix becomes
\beq
M_{\nu_R} \propto  \left
(\begin{array}{ccc}
\l_1^2\over  x & 0 & 0 \\
0 & \l_2^2\over  2 y & \l_2\over   2 y \\
0 & \l_2\over  2 y & 1\over  2 y
\end{array}
\right)
\eeq
which has some similarities with textures displayed in Table 1,
but is not identical.
This texture shows that, for large hierarchies in
$m_{eff}$ and an almost
diagonal 
$M_{\nu}^D$, the (23) mixing in 
$M_{\nu_R}$ scales
as $\lambda_2$.
This is consistent with what we found in
Figs. 3 and 4. we recall that
large hierarchies in $m_{eff}$ are
described by the limit $R \ra 1$. We see in
Fig. 4 that, for small Dirac 
hierarchies 
and negligible Dirac
mixing angle $\vartheta$, the angle
$\phi$ that describes (23) mixing in
$M_{\nu_R}$ has intermediate
values. However, as the Dirac hierarchies become
large, $\phi$ becomes very small,
as is indicated in
Fig.~3.

If, however, we take the Dirac mass matrix
to have maximal (23) mixing, the general texture
(\ref{twobytwo}) becomes
\beq
M_{\nu_R} \propto  \left
(\begin{array}{ccc}
\l_1^2\over  x & 0 & 0 \\
0 & (\l_2^2+y) \over  2 y & 
(-\l_2^2+y) \over  2 y  \\
0 & (-\l_2^2+y) \over  2 y  &
(\l_2^2+y) \over  2 y 
\end{array}
\right)
\label{textone}
\eeq
and clearly its form 
depends on the relative magnitudes
of $\lambda_{i},x$ and $y$. In the specific case where
the $m_{eff}$ hierarchy is much greater than the 
neutrino Dirac hierarchy: 
$\lambda_2 \gg y$, we
obtain the texture
\beq
M_{\nu_R} \propto {\rm Diag} 
{\left ( {\l_1^2\over  x}, {\l_2^2 \over 2y} {\left (
\begin{array}{cc}
1 & -1 \\
-1 & 1 \\
\end{array}
\right)}
\right)}
\label{texttwo}
\eeq
whereas when the $m_{eff}$ hierarchy is smaller: 
$\lambda_2 \ll y$ we find
\beq
M_{\nu_R} \propto {\rm Diag} {\left( {\l_1^2\over  x}, 
{1 \over 2} {\left(
\begin{array}{cc} 
1 & 1 \\
1 & 1 \\
\end{array}
\right)}
\right)}
\label{textthree}
\eeq
We note that (\ref{yminusone},\ref{texttwo},\ref{textthree})
span all but one of the possibilities for the $2 \times 2$
submatrix with indices (2,3).

We can again compare these solutions with the
results that we presented in Figs. 3 and 4,
in the region where the (23) Dirac mixing angle
$\vartheta$ becomes maximal. We see in
Figs.~3 and 4 that,
independently from the Dirac and
the $m_{eff}$ mass hierarchies,
as $\vartheta$ increases, so
does the required mixing in
$M_{\nu_R}$.
Moreover, for small
mass differences in $m_{eff}$,
the solution corresponds to
the last two examples of Table 1,
which indicate exactly
this effect.

\subsection{Mismatched Mixing}

A different structure arises when 
there is more than one
mixing angle in $m_{eff}$, or
there is a large
Dirac mixing angle that involves
different generations from those of the light Majorana matrix.
This happens, for example, when the atmospheric problem is solved
by $\nu_{\mu}\ra \nu_{\tau}$ oscillations whilst the Dirac mass
matrix is related to the quark mass matrix, with Cabibbo mixing
between the first and second
generations. The structure of the Majorana matrix
becomes more complicated for this {\it mismatched mixing}. 

In the case that $m_{eff}$ has two
large angles, the textures are
of course more complicated than in the previous subsection.
To see this, note that for an almost-diagonal
Dirac mass matrix, the wanted form
of the heavy Majorana mass matrix
for $y=-1$ 
becomes
\beq
M_{\nu_R} \propto  \left
(\begin{array}{ccc}
\l_1^2\over  2 x & -\l_1 \l_2\over  2 x  & \l_1 \over \sqrt{2} \\
-\l_1 \l_2\over  2 x & 
\l_2^2\over  2 x & \l_2\over  \sqrt{2} \\
\l_1 \over \sqrt{2} & \l_2\over  \sqrt{2} &  0
\end{array}
\right)
\eeq
and for $y \ll 1$ becomes
\beq
M_{\nu_R} \propto \left
(\begin{array}{ccc}
\l_1^2\over  2 x & -\l_1 \l_2\over  2 x  & 
-\l_1\over  2 \sqrt{2} y \\
-\l_1 \l_2\over  2 x & 
\l_2^2\over  2 x & -\l_2\over  2 \sqrt{2} y \\
-\l_1\over  2 \sqrt{2} y & -\l_2\over  2 \sqrt{2} y
 &  1\over  2 y
\end{array}
\right)
\eeq
In such a case, large Dirac hierarchies 
(and in particular 
$\lambda_1 \ll \lambda_2$),
effectively decouple the light entry of the
heavy Majorana mass matrix from the
heavier ones. 

This is no longer true, however, if
the (12) mixing angle in the
Dirac mass matrix becomes $\pi/4$,
and we still have two large mixing
angle in $m_{eff}$.

In this case
\bea
M_{\nu_R} \propto  \left
  (\begin{array}{ccc}
    2 \l_1^2 y + \l_2^2 x (1+y)\over  4xy &
    -2 \l_1^2 y + \l_2^2 x (1+y)\over  4xy &
    \l_2(-1+y)\over  2\sqrt{2} y \\
    -2 \l_1^2 y + \l_2^2 x (1+y)\over  4xy &
    2 \l_1^2 y + \l_2^2 x (1+y)\over  4xy &
    \l_2(-1+y)\over  2\sqrt{2} y \\
    \l_2(-1+y)\over  2\sqrt{2} y &
    \l_2(-1+y)\over  2\sqrt{2} y &
    (1+y)\over  2 y
  \end{array}
\right) \nonumber \newline
\hspace*{-0.2 cm}
\eea
which, in the limit of a large $m_{eff}$ hierarchy:
$y \ll 1$ and 
$\l_2^2 x \gg \l_1^2 y$
gives the texture
\bea
M_{\nu_R} \propto  \frac{1}{2y}
\times  \left
  (\begin{array}{ccc}
\l_2^2\over  2 & \l_2^2\over  2 & -\l_2\over  \sqrt{2}  \\ 
\l_2^2\over  2 & \l_2^2\over  2 & -\l_2\over  \sqrt{2}  \\ 
-\l_2\over  \sqrt{2}  & -\l_2\over  \sqrt{2}  & 1
  \end{array}
\right) 
\eea
Alternatively, if both
the (12) and (23) Dirac mixing angles are  maximal,
\beq
M_{\nu_R} \propto \left
  (\begin{array}{ccc}
    \frac{1}{4} \left (
      1+ {2 \l_1^2 \over  x} + {\l_2^2 \over  y }
    \right ) &
    \frac{1}{4} \left (
      1- {2 \l_1^2 \over  x} + {\l_2^2 \over  y }
    \right ) &
    (-\l_2^2 + y ) \over   2 \sqrt{2} y \\
    \frac{1}{4} \left (
      1-{2 \l_1^2 \over  x} + {\l_2^2 \over  y}
    \right ) &
    \frac{1}{4} \left (
      1+ {2 \l_1^2 \over  x} + {\l_2^2 \over  y }
    \right ) &
{    (-\l_2^2 + y ) \over   2 \sqrt{2} y} \\
 {   (-\l_2^2 + y ) \over   2 \sqrt{2} y} &
{    (-\l_2^2 + y ) \over   2 \sqrt{2} y} &
    {(\l_2^2 + y ) \over   2 y}
  \end{array}
\right)
\eeq
Once again, the exact form of the texture
depends on the relative mass hierarchies
in the various neutrino sectors.
As an example, in the double limit $y \gg \l_2^2$
and $x \gg 2 \l_1^2$ of the hierarchy factors,
which seems natural because Dirac masses exhibit
large hierarchies in many models,
we obtain:
\bea
M_{\nu_R} \propto \frac{1}{2}
\times  \left
  (\begin{array}{ccc}
{1 \over 2} & {1 \over 2} & {1 \over \sqrt{2}} \\
{1 \over 2} & {1 \over 2} & {1 \over \sqrt{2}} \\
{1 \over \sqrt{2}} & {1 \over \sqrt{2}}   & 1 
  \end{array}
\right) \nonumber 
\eea
Examining the above cases, we see, as expected, that
no simple-minded $2 \times 2$ substructure emerges.
Moreover, the precise way in which the various entries in the
full $3 \times 3$ matrix are filled depends on details of the
mass hierarchies studied.

\subsection{Related Neutrino Dirac and Quark Mixing}

Finally, we examine
more explicitly an example where
the atmospheric problem is solved
by $\nu_{\mu}\ra \nu_{\tau}$ oscillations whilst the neutrino Dirac mass
matrix is related to the $u$-quark mass matrix, with its CKM mixing.
Such mixing in the Dirac sector
arises naturally in some unified models, 
such as those related to $SO(10)$, and may in general be 
significantly different from the pattern of the heavy
Majorana mass matrix.

If $\tilde{s}_{1}, \tilde{c}_{1}$ refer here to Cabbibo mixing,
$\theta$ is the (23) neutrino mixing angle, and we neglect possible
(12) and (13) neutrino mixing,
the resulting
heavy Majorana form is
\bea
M_{11}&= &(m_u \tilde{c}_1^2 + m_c \tilde{s}_1^2)^2/m_1
+ (m_c-m_u)^2
( \cos^2\theta/m_2+\sin^2\theta/m_3)
 (\tilde{s}_1\tilde{c}_1)^2 \nonumber\\
M_{12}& =& (m_c-m_u) 
\tilde{c}_1\tilde{s}_1 \left [
 (m_u \tilde{c}_1^2 + m_c \tilde{s}_1^2)/m_1 +
\right.
\nonumber\\
          &+&\left.
(m_c \tilde{c}_1^2 + m_u \tilde{s}_1^2 )
( \cos^2\theta/m_2+\sin^2\theta/m_3) \right ] \nonumber \\
M_{13} & = &
 (m_2-m_3) m_t  (m_c-m_u) \tilde{s}_1 \tilde{c}_1
\sin(2 \theta) / (2 m_2 m_3) \nonumber \\
M_{22}& =& 
(m_c - m_u)^2 (\tilde{c}_1 \tilde{s}_1)^2 / m_1 +
(m_c \tilde{c}_1^2 + m_u \tilde{s}_1^2 )^2 
( \cos^2\theta/m_2+\sin^2\theta/m_3) \nonumber \\
M_{23}&= & (m_2 - m_3) m_t \sin(2\theta) 
(m_c \tilde{c}_1^2 +m_u \tilde{s}_1^2) / (2 m_2 m_3) \nonumber\\       
M_{33}&=& 
m_t^2 ( \cos^2\theta/m_3+\sin^2\theta/m_2)
\label{George}
\eea
To get an idea of the heavy neutrino textures that arises in this case,
we present two representative numerical examples,
for small and large neutrino mass
hierarchies. We take the following
conditions: a Cabibbo angle $\sim 12^0$, 
a near maximal (23)
neutrino mixing angle $\theta \sim 44.5^0$,
negligible (12)  and (13) mixings in $m_{eff}$,
$m_u = $ 5 MeV, $m_c = $ 1.4 GeV
and $m_t = $ 174 GeV. 

In the first example,
we consider light neutrino masses with the condition
$m_1:m_2:m_3 = 0.01:0.1:1.0$,$m_3 \approx 0.1$ eV,
In this case,
the numerical matrix is of the form:
\bea
M_{\nu_R} \propto
\left (
\begin{array}{ccc}
8.8 \times 10^9 & 4.0 \times 10^{10} &
      -2.2 \times 10^{12}\\
 4.0 \times 10^{10} &
     1.8 \times 10^{11} & - 10^{13}\\
      -2.2  \times 10^{12}& - 10^{13}&
      1.6 \times 10^{15}
\end{array}
\right )
\label{num1}
\eea
In terms of an expansion
parameter $\eps\approx 0.42$, $M_{\nu_R}$ can be parametrized as
follows:
\bea
M_{\nu_R}\propto
\left (
\begin{array}{ccc} 
\eps^{14}  &   \eps^{12}  &   -\eps^8\\
\eps^{12}  &   \eps^{10}  &   -\eps^6\\
-\eps^8  &    -\eps^6   &    1
\end{array}
\right )  
\label{nume1}
\eea 
On the other hand, for $m_3 = -m_2=m_1 \approx 1$ eV,
and the same parameters for $m_{eff}$
and $m_{\nu}^D$,
the heavy Majorana mass is numerically
\bea
M_{\nu_R} \propto
\left (
\begin{array}{ccc}
2.9 \times 10^6& 1.2 \times 10^7 &
      4.9 \times 10^{10}\\
 1.2\times 10^7 & 4.9\times 10^7& 2.3\times 10^{11}\\
      4.9 \times 10^{10}& 2.3 \times 10^{11}&
      5.3\times 10^{11}
\end{array}
\right )
\label{num2}
\eea
In terms of the same expansion
parameter $\eps\approx 0.42$, $M_{\nu_R}$ becomes:
\bea
M_{\nu_R}\propto
\left (
\begin{array}{ccc} 
\eps^{14}&\eps^{12}&\eps^3\\
\eps^{12}&\eps^{10}&\eps\\
\eps^3&\eps&1
\end{array}
\right )  
\label{nume2}
\eea 
where again ${\cal O}(1)$ coefficients are ignored. 

Comparing these two examples, we notice the change of the
required form of the heavy Majorana mass matrix
for large mixing in $m_{eff}$, as
we pass from large to small neutrino mass hierarchies.
Note in particular the increase in both the (23) 
and  the (13) mixing angles of
$M_{\nu_R}$ as we
pass from large to small hierarchies in
$m_{eff}$. It will be interesting later to
compare the qualitative features of
the two structures (\ref{nume1},\ref{nume2})
with the predictions of a specific flipped $SU(5)$ model.

The above cases exemplify
textures that lead to explanations of the
Super-Kamiokande data, in analogy with the
$2 \times 2$ cases that we discussed
in Section 3. For a given Dirac mass matrix,
the viable forms of the heavy Majorana masses
are quite constrained. As we discuss now, these phenomenological
textures may constrain severely the types of
flavour symmetries that could lead to
large neutrino mixing in realistic models.

%%%%%%%%%%%%%%%%%%%%%%%%%    

\section{Comments on Neutrino Textures and Flavour Symmetries}

  In many models, the structure of the fermion
  mass matrices, including those of the neutrinos,
  is dictated by family symmetries, for which
  the simplest possibility is
  a single Abelian $U(1)$ symmetry.
  The structure of the matrices is
  controlled by the flavour charges of the various
  fields: if an operator
  has zero total charge, then it is allowed
  in the low-energy Lagrangian.
  Usually, one assumes that
  the light Higgs charges are such
  that only
  the (3,3) renormalisable Yukawa coupling to $H_{2}$
  is allowed, plus that to $H_1$ in the case of large tan$\beta$. 
  The remaining entries are generated via the
  spontaneous breaking of the $U(1)$  symmetry, by the
  vev's of singlet fields
  $\langle\theta\rangle, \langle\bar{\theta}\rangle$, 
  with $U(1)$ charges $\pm 1$
  in the simplest case. Here we make just a few remarks about such models.

  The first step in describing neutrino
  masses is to determine the Dirac and heavy Majorana
  mass matrices. The simplest case arises when
  we add three generations
  of right-handed neutrinos,
  leading to predictions for light neutrino
  masses through the see-saw mechanism as above. In such a
  model, $SU(2)$ invariance fixes the charges
  of the left-handed neutrino states to be the
  same as those of the charged leptons. Then,
  if one imposes a left-right symmetry the
  charges of the right-handed neutrinos are also fixed.
  In the case of asymmetric mass matrices, 
  there is more freedom in the choice of the charges,
  but, in specific models we
  discuss later, the $U(1)$ charges of the 
  various fields can be correlated.

  The Majorana mass terms for the right-handed neutrinos
  arise from contributions of the form
  $\nu_R^i. \nu_R^j {\rm (singlet)}_1^m {\rm (singlet)}_2^n ...$,
  where the ${\rm (singlet)}_i$ stand for $SU(3) \otimes
  SU(2) \otimes U(1)$-invariant scalar fields.
  The various choices for the charges of
  the singlet fields lead to a variety of
  possible forms for the Majorana mass,
  which recur in richer models where 
  more than one type of singlet field can be present.
  The implications of such models will be manifest in a subsequent section,
  when we discuss a specific model, namely string-derived flipped $SU(5)$.
  For the moment, let us initially
  assume the existence of a field $\Sigma$ with
  charge opposite to that of some given combination
  $\nu^R_i \nu^R_j$. This automatically
  allows the  $(i,j)$ entry of the
  heavy Majorana mass matrix to be of order unity, whilst the rest of its
  entries are generated by non-renormalisable
  contributions and are therefore suppressed.
  If $i=j$, the largest entry will be on the diagonal,
  which is analogous to the generic form usually studied
  for Dirac mass matrices. However,
  if $i\neq j$ an effective submatrix of the form
  \bea
  \left (
    \begin{array}{cc}
      0 & 1 \\
      1 & 0 
    \end{array}
  \right ) \nonumber 
  \eea
  appears, suggesting that large mixing may
  be generic. 
This is the case in particular because it is difficult to generate
additional large entries if there is only
one singlet field $\Sigma$, in addition to
$\theta,\bar{\theta}$. However, extra terms
can be generated
if additional singlets are available~\cite{ver}. 
In each case, the dominant elements of the mass matrix
will be determined by the vev's of the singlet fields,
and the order of the non-renormalisable operators~\footnote{Besides the
relative magnitudes of the neutrino masses,
their absolute magnitudes also depend on
the vev's of the singlet fields. The requirement of obtaining
realistic mass scales for neutrino physics 
can be used to constrain the possible
flat directions in specific models.}.

In a previous section, we examined the possible 
forms of phenomenological textures that may lead
to large mixing, and we now illustrate their use to constrain
theoretical models with flavour symmetries.
Suppose that we have a model with a single 
$U(1)$ symmetry, under which quarks and leptons
have the same charge~\cite{IR}. Then, for the 
$2 \times 2$ quark Dirac mass matrices  one has 
the forms
\beq
m^D_u =
\left (
\begin{array}{cc}
\epsilon^{2p+h_2} & \epsilon^{p+q+h_2} \\
\epsilon^{p+q+h_2} & \epsilon^{2q+h_2}
\end{array}
\right),
m^D_d =
\left (
\begin{array}{cc}
\bar{\epsilon}^{2p+h_1} & \bar{\epsilon}^{p+q+h_1} \\
\bar{\epsilon}^{p+q+h_1} & \bar{\epsilon}^{2q+h_1}
\end{array}
\right)
\end{equation}
where $p,q$ are the charges of the second and third 
generation quarks, and $h_{1,2}$ the charges of the 
Higgs fields.
Obtaining the correct mass hierarchies
and  $V_{CKM}$ mixing automatically
implies that the up- and down-quark
mass matrices have similar structures,
with the (1,2) and (2,1) entries
larger than the (2,2) ones.
In this case, a large mixing 
angle in the heavy Majorana mass matrix may
not get communicated to $m_{eff}$ \cite{DLLRS},
and the large mixing in
$m_{eff}$ would have to arise mainly
from $m_{\nu}^D$  and the charged lepton
sectors \cite{LLR}.

This analysis gets modified if: \\
\noindent
1) Neutrinos and up-type quarks of
the same generation do not belong to the
same multiplets of the gauge group. Then 
we can have diagonal neutrino mass matrices,
and non-diagonal quark ones. \\
\noindent
If, however, we require similar neutrino and quark
structures, and still want to carry
large mixing in $m_{\nu_R}$ over to
$m_{eff}$, we have alternatives, of which
the first is the following. \\
\noindent
2) Asymmetric mass matrices with
different charges for up and down
quarks yield different structures for the mass
matrices. \\
\noindent
In this case, the $V_{CKM}$ mixing may arise entirely
from one sector, e.g., the down quarks, whilst in the
up sector we may have an almost diagonal form, with the
only significant requirement being that of getting the correct
$m_c/m_t$
ratio. We note that in realistic GUT models, such as
the one we discuss below, the
Dirac mass matrices are indeed expected to be asymmetric,
since the up and down quarks are assigned
to different representations of the GUT group.
Moreover, even in models where we combine 
flavour symmetries with GUTs where 
particles of the
same generation belong to the same multiplets,
the existence of different Clebsch-Gordan coefficients
can lead to additional zeroes beyond those
of the flavour symmetry, and thus to
asymmetric textures, even if we start with symmetric charges. \\
\noindent
3) Alternatively, one may have symmetric mass matrices, but
the up and down matrices may have different structures of
zero elements. \\
\noindent
This can again arise either because of 
zero Clebsch-Gordan coefficients,
or in the presence of additional residual symmetries~\cite{ver}. In this
case, we can again
obtain the correct $V_{CKM}$ mixing entirely
from one sector, and have almost-diagonal forms
for the up-quark and Dirac-neutrino masses.

\section{Neutrino Mixing in a Realistic Flipped $SU(5)$ Model}

Let us now look at a specific example of the structure
generated by $U(1)$ symmetries, namely the 
Ansatz made in \cite{ELLN} in the context of a `realistic'
flipped $SU(5)$ model derived from string, which is reviewed
in the Appendix~\footnote{For previous studies of fermion mass matrices in
this model, see~\cite{fermaU1,neutrinoFL}. Neutrino masses
have been  studied in~\cite{DIML,neutrinoFL}.}.
This model contains 
many singlet fields, and the mass matrices depend on the subset of these
that get non-zero vev's, which in turn depends on the 
choice of flat direction in the effective potential, which is
ambiguous, so far.

\subsection{Charged-Lepton Masses and Mixing}

In previous sections, we worked in a field basis that was
diagonal for the mass eigenstates of the charged leptons.
In the context of the flipped $SU(5)$ model, this has to be
identified relative to the string states listed in the
Appendix, which requires a discussion of the charged-lepton mass
matrix. The importance of this discussion lies in the possibility
that there might be additional mixing coming from this
sector. In this connection, we
recall that the mixing angles of
relevance to experiment
are the combinations given by
\beq
V_{\nu} = V^{m \dagger}_{\nu} V^m_{\ell_L}
\label{bothmix}
\eeq
where the symbols $V^m_{\nu}, \ell_L$ denote the rotation
matrices for neutrinos and left-handed charged leptons,
respectively, required to diagonalize their mass matrices.

The candidate terms for charged-lepton mass terms
at the third-order level are
\begin{equation}
\bar{f}_1 \ell^c_1 h_1, \; \;
\bar{f}_2 \ell^c_2 h_2, \; \;
\bar{f}_5 \ell^c_5 h_2.
\end{equation}
where, here and later, we do not display factors of the
gauge coupling.
The first term generates the $\tau$ mass, but
since the last two are proportional to the same Higgs
$h_2$, they cannot yield a mass hierarchy. We therefore
assume that the vev of the effective light Higgs has only
a small component in the $h_2$ direction, as also
assumed in~\cite{ELLN}. Thus, 
in a first approximation we assign
$\ell^c_1$ and the charged component of ${\bar f}_1$
to the $\tau$, and the corresponding $\ell^c_{2,5}, {\bar f}_{2,5}$
to the $e, \mu$, with the precise flavour assignments of the latter
to be discussed below.

Assuming a very small vev for $h_2$, the next
candidate mass terms appear at fifth order~\footnote{Here and 
subsequently, higher-order interactions should always
be understood to be scaled by the appropriate inverse power of
some relevant dimensional scale $M_s$. We expect this to be ${\cal O}
(10^{18})$ GeV in conventional string theory, but it might be as low as
$\sim 10^{16}$ GeV in $M$ theory. The vev's we quote later for
singlet fields are likewise in units of $M_s$.}~\cite{DIML}:
\begin{equation}
\bar{f}_2 \ell^c_2 h_1 (\bar{\phi}_i^2 +
\bar{\phi}^+\bar{\phi}^-), \;\; \;\;
\bar{f}_5 \ell^c_5 h_1 (\bar{\phi}_{1,4}^2 +
\bar{\phi}^+\bar{\phi}^-)
\end{equation}
Among the fields in parentheses, previous analyses suggest (see
the Appendix) that ${\bar \phi}_{1,2}$ and ${\bar \phi}^-$ have
zero vev's. Therefore the possible mass terms are
\begin{equation}
\bar{f}_2 \ell^c_2 h_1 \bar{\phi}_{3,4}^2, \;\; \;\;
\bar{f}_5 \ell^c_5 h_1 \bar{\phi}_{4}^2
\end{equation}
It is apparent that, in order to obtain a hierarchy: $m_{\mu} \gg m_e$,
we must assume that either ${\bar \phi}_3^2 \gg {\bar \phi}_4^2$ or
the inverse. As we argue later on the basis of the $u$-quark masses
and mixing that ${\bar \phi}_4 \ll 1$, we assume that ${\bar \phi}_3
\gg {\bar \phi}_4$.

Continuing to seventh order, we find the term:
\begin{equation}
\bar{f}_5 \ell^c_2 h_1 \Delta_2 \Delta_5 
(\bar{\phi}_i)^2
\label{leptonseven}
\end{equation}
but, to this order, we still find no term mixing
${\bar f}_1, \ell^c_1$ with the other lepton fields.
As mentioned in the previous
paragraph, we assume that
${\bar \phi}_3 \gg {\bar \phi}_4, {\bar \phi}_{1,2} = 0$.
The charged-lepton mass-mixing problem can therefore
be reduced to the following $2 \times 2$ matrix
in the ${\bar f}_{5,2},\ell^c_{5,2}$
basis:
\beq
m_{\ell} (2 \times 2) \propto \left (
\begin{array}{cc}
{\bar \phi}_4^2 & \Delta_2 \Delta_5 {\bar \phi}^2_3 \\
0 & {\bar \phi}^2_{3}
\end{array}
\right )
\label{leptontwo}
\eeq
where, again in view of the $u$-quark mass matrix discussed
below,
we believe that $\Delta_2 \Delta_5$ is not small.
Since ${\bar \phi}_3 \gg {\bar \phi}_4$, 
we assign the charged leptons to the
eigenvectors of (\ref{leptontwo}) as follows:
$(e^c, \mu^c) = (\ell^c_5, \ell^c_2)$ and $(e_L,
\mu_L) = ({\bar f}_5 - {\cal O}(\Delta_2 \Delta_5)
{\bar f}_2, {\bar f}_2 + {\cal O}(\Delta_2
\Delta_5){\bar f}_5)$, 
with the ratio of mass eigenvalues
\beq
\frac{m_{\mu}}{m_e} \sim \frac{m_{\ell_1}} { m_{\ell_2}} \sim \frac{{\bar
\phi}_3^2}
{{\bar \phi}^2_{4}}
\label{leptoneigen}
\eeq
Thus we see explicitly that we can arrange
a hierarchy $m_{\mu} \gg m_e$, at the
price of a potentially
large mixing angle among the left-handed charged leptons: 
$V^m_{\ell_L}(12) = {\cal O} (\Delta_2 \Delta_5)$.
This would lead us naively to expect correspondingly large 
$\nu_e - \nu_{\mu}$ mixing, unless there
is some cancellation with $V^m_{\nu}$ in (\ref{bothmix}).

\subsection{Dirac Neutrino Masses}

Even with a given choice of a flat direction,
the neutrino mass matrix that arises from
the string model is rather complicated,
because one must consider light Majorana, Dirac and heavy Majorana
mass matrices. The first of these could arise from
direct effective operators involving
two left-handed neutrinos, two light Higgs doublets, and singlet fields.
However, we find no candidates for such terms up to fifth
order, and shall not discuss them further here.
As for the Dirac mass matrix, since the neutrino flavours
are in the same representations as the $u$-type quarks,
with the left-handed neutrinos
belonging to the representations $\bar{f}_{1,2,5}$, whilst the
right-handed neutrinos naively belong to the decuplets $F_{2,3,4}$,
one would naively expect the relation
\bea
m_{\nu}^D = (m_u)^{\dagger}
\label{nuup}
\eea
However, one should also not forget that 
there may be Dirac mass couplings of light neutrinos
to singlet states not included among the
$F_{2,3,4}$, and that these fields may also mix with the
singlets via Majorana mass terms, possibilities
that will play important r\^oles later.

At third order, we find the following contribution to
the Dirac neutrino mass matrix, which corresponds to the
dominant contribution to $m_t$:
\begin{eqnarray}
F_4\bar{f}_5 \bar{h}_{45}
\label{diracthree}
\end{eqnarray}
%In addition, there is a term $\frac{1}{\sqrt{2}}\bar{F}_5 F_4\phi_3$
%that we do not discuss further here.
Progressing up to sixth order, the following additional terms appear:
\begin{eqnarray}
{}F_2\bar{f}_2 \bar{h}_{45}\bar{\phi}_4, &
{}F_4\bar{f}_2\bar{h}_{45}\Delta_2\Delta_5\\
{}F_2\bar{f}_5 \bar{h}_{45}\Delta_2\Delta_5\bar\phi_4,&
{}F_3\bar{f}_5\bar{h}_{h45}\Delta_3\Delta_5\bar{\phi}_3
\label{Diracsix}
\end{eqnarray}
We observe that the Dirac matrix 
again leaves the $\nu_1$ component of ${\bar f}_1$
essentially decoupled from the other light neutrinos,
up to sixth order. The most important mixing effects
are therefore expected to take place between $\bar{f}_2$ and 
$\bar{f}_5$, and
the problem can be reduced, in a first approximation,
to considering only two neutrino species.
This is equivalent to the $2 \times 2$ mixing
matrix for the two heaviest quark generations:
$m_{u} (2 \times 2) = m^{D^\dagger}_{\nu} (2 \times 2)$,
and some indications on the values of the vev's
appearing in (\ref{Diracsix}) may be obtained from
the experimental values of $m_c/m_t$ and the
$V_{CKM}$ parameters.

The $2 \times 2$ part of the
up-quark mass matrix for the two heavier generations
is of the following form~\cite{ELLN} in the $F_2, F_4$, ${\bar f}_2, {\bar
f}_5$ basis:
\begin{eqnarray}
m_{u} (2 \times 2) = m^{D^\dagger}_{\nu} (2 \times 2) =
\left ( 
\begin{array}{cc}
 \bar{\phi}_4 & \Delta_2 \Delta_5 \bar{\phi}_4 \\ 
 \Delta_2 \Delta_5 & 1
\end{array}
\right ) \; \lambda_t(M_{GUT})\langle\bar{h}_{45}\rangle ,
\label{twomasses}
\end{eqnarray}
This implies that the (23) $u_L$ mixing angle, which contributes to
$V_{CKM}$, is given by $\theta^{u_L}_{(23)} =
\Delta_2 \Delta_5 \bar{\phi}_4$,
whilst the (23) $u_R$ mixing angle is
$\theta^{u_R}_{(23}) = \Delta_2 \Delta_5$.
The corresponding mass eigenvalues are:
\begin{equation}
m_{u}^{1,2} \approx
\frac{1}{2} \left (
1+{\bar \phi}_4 \pm \sqrt{1-2{\bar \phi}_4 + 4 (\Delta_2 \Delta_5)^2 {\bar
\phi}_4
+{\bar \phi}_4^2} \right)
\end{equation}
so we see that the heavier eigenvalue is almost unity,
whilst the lighter is suppressed if ${\bar \phi}_4 \ll 1$:
\beq
{m_c \over m_t} \sim {\bar \phi}_4 \times {\cal O}(1)
\label{mcovermt}
\eeq
One
should not be too concerned at this stage about the compatibility of this
equation with (\ref{leptoneigen}), since unknown numerical
factors remain to be calculated.
More information about the vev's of the fields is provided by
the (23) element of $V_{CKM}$. This also receives a contribution from
the (23) element of the down-quark mass matrix,
which was also found~\cite{ELLN} to be of order
$\Delta_2 \Delta_5 \bar{\phi}_4$.
Up to constants of order unity, which we do not keep track of in
our analysis of mass matrices, we conclude that
\beq
\Delta_2 \Delta_5 \bar{\phi}_4 \approx 0.044
\label{onecombination}
\eeq
We see from (\ref{mcovermt}) that having ${\bar \phi}_4$ 
large and $\Delta_2 \Delta_5$ small will not give
acceptable solutions. However,
the choice of large $\Delta_2 \Delta_5$ and smaller ${\bar \phi}_4$
does lead to acceptable solutions. 
For example, fixing ${\bar \phi}_4 \approx 0.044 / \Delta_2 \Delta_5$, we
find for $\Delta_2 \Delta_5 \approx 0.8$ that $m_c / m_t = 0.018$, whilst
for 
$\Delta_2 \Delta_5 \approx 0.9$ we find $m_c / m_t = 0.008$.
However, we should also note that the values of the
acceptable field vev's are sensitive to the presence of
order unity coefficients.
In particular,
$\Delta_2 \Delta_5$ can become smaller.
For example, if the off-diagonal elements in (\ref{twomasses})
happen to be multiplied by factors of two,
we find for $\Delta_2 \Delta_5 = 0.47$:
$m_c / m_t = 0.009$, and 
for $\Delta_2 \Delta_5 = 0.53$:
$m_c / m_t = -0.009$, 
whilst for $\Delta_2 \Delta_5 = 0.5$:
$m_c / m_t \approx 0$. 

This is why we assumed that $\Delta_2 \Delta_5$ is large
and ${\bar \phi}_4 \ll 1$ in our earlier analysis of the
charged-lepton mass matrix, which then required ${\bar \phi}_3 = {\cal
O}(1)$. Analysis of the (13) entry in
$V_{CKM}$, which is ${\cal O}(\Delta_3 \Delta_5 {\bar \phi}_3)$,
might then lead one to suspect that $\Delta_3 \ll 1$. However, as
can be seen from~\cite{ELLN}, this would lead to too small a value
for the Cabibbo angle. In fact, it is not necessary that $\Delta_3 \ll 1$,
since (unlike the (12) entry)
the (13) entry in $V_{CKM}$ results from a difference between
two terms of the same order originating from $u$- and $d$-quark mixing,
and there could be a cancellation between them,
depending on the precise numerical coefficients.

We have omitted from the above discussion the last term in
(\ref{Diracsix}), which includes factors of $\Delta_3$ and
${\bar \phi}_3$. We have no strong reason
to neglect this term, except for the fact that it is of
sixth order. Nevertheless, we assume for simplicity that
this and other mixing with $F_3$ can be neglected as a first
approximation.
Absent from the above discussion has been any Dirac neutrino
mass term involving
${\bar f}_1$. There is no such coupling to any of the $F_{2,3,4}$
up to sixth order, but
there is such a coupling to $\phi_1$ in fourth order:
\begin{equation}
F_1\bar{f}_1\bar{h}_{45}\phi_1,
\label{thirdheavy}
\end{equation}
which may lead to
mixing between the $\nu_1$ component of ${\bar f}_1$ and the singlet
$\phi_1$, if $F_1$ develops a vev~\cite{TR}.
Since the term (\ref{thirdheavy}) is only fourth order,
we consider $\phi_1$ as the best candidate for the
third $\nu_R$ state, rather than one of the $F_i$. 

This example serves to warn us
that the expected relation (\ref{nuup}) may be too naive, the reason
being that the $u$ quark is so light that some other effect, such as
mixing with additional heavy singlet states, may be important.

\subsection{Heavy Majorana Masses}

We now discuss the heavy Majorana mass matrix for the
fields $F_2, F_4$, which we parametrize as:
\begin{equation}
\left(
\begin{array}{cc}
M & M'\\
M' & M^{\prime\prime}
\end{array}
\right)
\label{heavymaj}
\end{equation}
As we now discuss, the heavy Majorana entries $M,M'$ and
$M''$ are expected to be generated from higher-order 
non-renormalizable terms. Their magnitudes
play crucial r\^oles in the mixing of the light neutrinos, as the
previous simple $2\times 2$ and $3\times 3$ phenomenological analyses has
shown. We find candidate terms for the $M,M'$ contributions at
seventh order. Up to this order,
a complete catalogue of the operators that could
generate heavy Majorana neutrino mass terms
involving the fields $F_2$ and $F_4$ is given by:
\bea
W_{NR} & = &  F_2  F_2   (\bar{F}_5  \bar{F}_5  \bar{\Phi}_2  \phi_1 +
\bar{F}_5  \bar{F}_5  \bar{\phi}_3  \phi_4  + 
\bar{F}_5  \bar{F}_5  \bar{\phi}_4  \phi_3  + \bar{F}_5  \bar{F}_5  \bar{\phi}_1  \phi_2 +
\nonumber \\
& &  
  \bar{F}_5  \bar{F}_5  \phi_{45}  \bar{\phi}_{45}  \Phi_4 +
  \bar{F}_5  \bar{F}_5  \phi^{-}  \bar{\phi}^{-}  \Phi_4 +
  \bar{F}_5  \bar{F}_5  \phi^{-}  \bar{\phi}^{+}  \Phi_4  +
  \bar{F}_5  \bar{F}_5  D_5  \bar{\phi}_3  D_4 + \nonumber \\ & &
  \bar{F}_5  \bar{F}_5  \bar{\Phi}_2  \phi_1  \Phi_1 + 
  \bar{F}_5  \bar{F}_5  \bar{\Phi}_2  \phi_1  \Phi_3 +
  \bar{F}_5  \bar{F}_5  \bar{\Phi}_2  \phi_2  \Phi_4 +
  \bar{F}_5  \bar{F}_5  \bar{\phi}_3  \phi_3  \Phi_4 +  \nonumber \\ & &
  \bar{F}_5  \bar{F}_5  \bar{\phi}_3  \phi_4  \Phi_1 +
  \bar{F}_5  \bar{F}_5  \bar{\phi}_3  \phi_4  \Phi_3 +
  \bar{F}_5  \bar{F}_5  \bar{\phi}_3  \phi_4  \Phi_5 +
  \bar{F}_5  \bar{F}_5  \bar{\phi}_4  \phi_3  \Phi_1 +\nonumber \\ & & 
  \bar{F}_5  \bar{F}_5  \bar{\phi}_4  \phi_3  \Phi_3 +
  \bar{F}_5  \bar{F}_5  \bar{\phi}_4  \phi_3  \Phi_5 +
  \bar{F}_5  \bar{F}_5  \bar{\phi}_4  \phi_4  \Phi_4 + 
  \bar{F}_5  \bar{F}_5  \bar{\phi}_1  \phi_1  \Phi_4 + \nonumber \\ & &
  \bar{F}_5  \bar{F}_5  \bar{\phi}_1  \phi_2  \Phi_1 +  
  \bar{F}_5  \bar{F}_5  \bar{\phi}_1  \phi_2  \Phi_3  ) + \nonumber \\ & &
 F_4  F_4  (
 \bar{F}_5  \bar{F}_5  \phi_1  \phi_2   +   \bar{F}_5  \bar{F}_5  \phi_3  \phi_4   +
\nonumber \\ & &
 \bar{F}_5  \bar{F}_5  \phi^{-}  \phi^{+}  \Phi_4 +
 \bar{F}_5  \bar{F}_5  D_5  \phi_3  D_4 +
 \bar{F}_5  \bar{F}_5  \phi_1  \phi_1  \Phi_4 +
 \bar{F}_5  \bar{F}_5  \phi_1  \phi_2  \Phi_1 + \nonumber \\ & & 
 \bar{F}_5  \bar{F}_5  \phi_1  \phi_2  \Phi_2 +
 \bar{F}_5  \bar{F}_5  \phi_1  \phi_2  \Phi_3 + 
 \bar{F}_5  \bar{F}_5  \phi_2  \phi_2  \Phi_4 +
 \bar{F}_5  \bar{F}_5  \phi_3  \phi_3  \Phi_4 + \nonumber \\ & &
 \bar{F}_5  \bar{F}_5  \phi_3  \phi_4  \Phi_1 +
 \bar{F}_5  \bar{F}_5  \phi_3  \phi_4  \Phi_2 + 
 \bar{F}_5  \bar{F}_5  \phi_3  \phi_4  \Phi_3 +
 \bar{F}_5  \bar{F}_5  \phi_3  \phi_4  \Phi_5 + \nonumber \\ & &
 \bar{F}_5  \bar{F}_5  \phi_4  \phi_4  \Phi_4 ) + \nonumber \\ & &
F_2 F_4 \bar{F}_5 \bar{F}_5 \Delta_2 \Delta_5 \phi_3
\label{wotalot}
\eea
Please note that we include
at this stage even some combinations
involving singlet fields which we had assumed in~\cite{ELLN} 
(see also the Appendix) to have zero vev's.  This is
done in order to develop a more general picture
of the types of terms that are allowed.
However, we have dropped combinations of the
type $D_i^2$, since such terms would
not allow for two light Higgses.

The only term in (\ref{wotalot}) that involves the combination
$F_2 F_4$ is
$F_2 F_4 \bar{F}_5 \bar{F}_5 \Delta_2 \Delta_5 \phi_3$.
Previously, in \cite{ELLN},
where we studied the implications 
of this model for the quark 
mass matrices, we assumed that
$\phi_3 = 0$. However, 
this restriction may be avoided~\cite{TR} by a different choice of
flat direction~\footnote{In this connection, it
is worth noting that there is more freedom in assigning non-zero vev's
to the various singlets if one allows for additional
phases, beyond those introduced in~\cite{ELLN}.
A modification of the pattern of vev's
would entail a modified discussion of the flatness conditions
at higher order, but a complete analysis goes beyond the
scope of this paper.}.
If we adopt the minimal modification of the
flat direction chosen in \cite{ELLN}
that allows for a non-zero vev for $\phi_3$,
none of the additional terms 
involving $F_4 F_4$ survives.
However, there is an
effective term $F_2 F_2 \bar{F}_5 \bar{F}_5 \bar{\phi}_4 \phi_3$
that provides $F_2 F_2$ mixing~\footnote{We recall that the
Higgs mass matrix mixes the
pentaplets $h_{1,2,3,45}$ and 
their conjugate fields, and needs to have two
massless combinations. Keeping the rest of the
field vev's as in \cite{ELLN},
the inclusion of a non-zero vev for $\phi_3$
gives a new 
contribution only when we include the
$f_4$ field, which also contains an electroweak doublet. A coupling
$h_{45} f_4 \bar{F}_5 \bar{\phi}_4^2 \phi_2 \phi_3$
is generated at seventh order.
However, there are still two massless
states left in the $4 \times 4$ space of
the $h_{i,ij},\bar{h}_{i,ij}$ fields.}.
We therefore conclude that, to seventh order, this model has:
\bea
M= \bar{F}_5 \bar{F}_5 \bar{\phi}_4 \phi_3,\;\;
M'=\bar{F}_5 \bar{F}_5 \Delta_2 \Delta_5\phi_3,\;\; M^{\prime\prime}=0
\label{gotsome}
\eea
Clearly, the form of the
heavy Majorana mass matrix depends on the
relative magnitudes of the vev's of  the 
$\Delta_2 \Delta_5$ and $\bar{\phi}_4$
field combinations, which we discussed earlier in connection
with the matrix $m_u = m^D_\nu$.

This does not complete our discussion of the heavy Majorana
mass matrix, since we should also discuss possible
mass terms involving $\phi_1$, our candidate for the third
$\nu_R$ state. To seventh order, the following are the only
such candidate Majorana mass terms we find:
\beq
\phi_1 F_4 {\bar F}_5 {\bar \phi}_{31} \phi_{31} {\bar \phi}_4 \phi_2
\ra M_{4\phi}, \;\;\;\;
\phi_1^2 \Delta_2 \Delta_5 {\bar \phi}_{23} T_2 T_5 \ra M_{\phi\phi}
\label{phimaj}
\eeq
The first of these mixes $\phi_1$ with $F_4$, and the latter is a diagonal
Majorana mass term. Combining these with (\ref{gotsome}), we find
the following $3 \times 3$ heavy Majorana mass matrix in the $F_2, F_4,
\phi_1$ basis:
\beq
\left (
\begin{array}{ccc}
M & M' & 0 \\
M' & 0 & M_{4\phi} \\
0 & M_{4\phi} & M_{\phi \phi}
\end{array}
\right )
\label{heavymaj2}
\eeq
Since all of these terms arise in seventh order, and the
vev's appearing in them are not very tightly constrained,
diagonalization of
the heavy Majorana mass matrix may well require large mixing
angles, but these cannot be predicted accurately. Nevertheless, it would
seem to be a general feature that the characteristic heavy Majorana
mass scale $M_N \ll M_s$, since all the entries in (\ref{heavymaj2})
are of high order, with several potentially small vev's. This makes
the appearance of one or more neutrino masses around $0.1$ eV quite
natural, as we discuss now.

\subsection{Neutrino Mass Textures in Flipped $SU(5)$}

As a preliminary to constructing the neutrino mass
matrices, we first recall the left-handed charged-lepton assignments
motivated earlier: $(e_L,\mu_L, \tau_L)$ = $({\bar f}_5 - 
{\cal O}(\Delta_2 \Delta_5) {\bar f}_2$,
${\bar f}_2 $+$ {\cal O}(\Delta_2 \Delta_5){\bar f}_5,
{\bar f}_1)$.
The weak-interaction eigenstates for the light neutrinos must have the
same assignments:
\bea
\nu_{e} \rightarrow \bar{f}_5 - {\cal O}(\Delta_2 \Delta_5) {\bar f}_2,
\;\;\;\;
\nu_{\mu} \rightarrow \bar{f}_2 + {\cal O}(\Delta_2 \Delta_5){\bar f}_5
\;\;\;\;
\nu_{\tau} \rightarrow \bar{f}_1 
\label{assignneutrinos}
\eea
However, it is convenient to work in the basis
$({\bar f}_5, {\bar f}_2, {\bar f}_1)$,
which is related to (\ref{assignneutrinos}) by the rotation
\beq
V^m_{\ell_L} = \left (
\begin{array}{ccc}
1-\frac{1}{2} (\Delta_2 \Delta_5)^2 & \Delta_2 \Delta_5 & 0 \\
-\Delta_2 \Delta_5  &  1-\frac{1}{2} (\Delta_2 \Delta_5)^2 & 0 \\
0 & 0 & 1
\end{array}
\right )
\label{veenum}
\eeq
As for the massive right-handed neutrinos,
the coupling (\ref{thirdheavy}) means that
$\nu_{\tau_R}$ has to be assigned to
$\phi_1$, since it is the only field to which
${\bar f}_1$ couples at a significant level.
In view of the couplings (\ref{diracthree},\ref{Diracsix}), we assign
$\nu_{\mu_R}$ to $F_4$ and
$\nu_{e_R}$ to $F_2$.

With these choices of bases, $m_{\nu}^D$ takes the form
\begin{equation} 
m_{\nu}^D = 
\left (
\begin{array}{ccc}
\Delta_2 \Delta_5 \bar{\phi}_4 & 1 & 0 \\
\bar{\phi}_4 & \Delta_2 \Delta_5 & 0 \\
0 & 0 & F_1
\end{array}
\right)
\label{Dira}
\end{equation}
whilst $M_{\nu_R}$ is given by
\begin{equation} 
M_{\nu_R} = 
\left (
\begin{array}{ccc}
\bar{F}_5 \bar{F}_5  \bar{\phi}_4 \phi_3 &
\bar{F}_5 \bar{F}_5  \Delta_2 \Delta_5 \phi_3 & 0 \\
\bar{F}_5 \bar{F}_5  \Delta_2 \Delta_5 \phi_3 &
0 & \bar{F}_5 \bar{\Phi}_{31}\Phi_{31} \bar{\phi}_4 \phi_2 \\
0 & \bar{F}_5 \bar{\Phi}_{31}\Phi_{31} \bar{\phi}_4 \phi_2 
& \Delta_2 \Delta_5 \bar{\Phi}_{23} T_2 T_5
\end{array}
\right)
\label{hema}
\end{equation}
The resulting $m_{eff}$ is given by (\ref{eq:meff}), and the
neutrino mixing angles in the weak-eigenstate basis
(\ref{assignneutrinos}) are given by (\ref{bothmix}).

Clearly, the forms of the mass matrices depend
on the various field vev's. For these, we have some
information from analysis of the flat directions and the rest
of the fermion masses, but there is still
some arbitrariness. For example, in the cases of the
decuplets that break the gauge group down to
the Standard Model, we know that the vev's 
should be 
$\approx M_{GUT}/M_{s}$. In weakly-coupled string constructions,
this ratio is $\approx 0.01$. However, the strong-coupling limit of
$M$ theory offers the possibility that the
GUT and the string scales can coincide, in which case the
vev's could be of order unity. 

What about the other fields?
The analysis of quark masses suggested that
$\Delta_2 \Delta_5$ should be of order unity,
whilst $\bar{\phi}_4$ should be suppressed.
The analysis of flat directions in~\cite{ELLN} indicate that
if $\bar{\phi}_3^2$ is large, as we have suggested in order
to get the correct $m_{e}/m_{\mu}$ ratio,
then $\bar{\Phi}_{31} \bar{\Phi}_{23}$ 
is also large. The flatness conditions~\cite{ELLN} 
relate $\bar{\Phi}_{31},\Phi_{31}$ and
$\phi_2$, and can be satisfied even if all
the vev's are large, as long as
$\bar{\Phi}_{31} \Phi_{31}$ and
$\bar{\Phi}_{23} \Phi_{23}$ are
not very close to unity.
Finally, we note that nothing yet fixes the value of
$T_2 T_5$.

Despite these uncertainties, the following features
of the mass matrices are apparent. (i) The heavy
Majorana matrix $M_{\nu_R}$ is likely to have many entries that may
be of comparable magnitudes. In particular, (ii) there are
potentially large off-diagonal entries that could yield large
$\nu_{\mu} - \nu_{\tau}$ and/or $\nu_{\mu} - \nu_{e}$ mixing.
(iii) The neutrino Dirac matrix is {\it not} equivalent to $m_u$,
and (iv) is also a potential source of large $\nu_{\mu} -
\nu_{e}$ mixing. We recall (v) that charged-lepton mixing is
potentially significant
and note that, in general, (vi)
the mass matrices
(\ref{Dira},\ref{hema})
correspond to the {\it mismatched mixing}
case of Section 5.
Finally, we recall (vii) that there is significant
mixing of candidate $\nu_R$ states with singlet fields.

A complete analysis of the available parameter
space goes beyond the scope of this paper, and would perhaps involve
placing more credence in the details of this model than it deserves.
Accordingly, we limit ourselves to some general comments on the
likelihood of mass degeneracies relative to hierarchies in $m_{eff}$,
and on the plausibility of large mixing in the 
$\nu_{\mu} - \nu_{\tau}$ and/or $\nu_{\mu} - \nu_{e}$ sectors.

To this end, we first consider the following simplified forms
for the matrices (\ref{leptontwo},\ref{twomasses},\ref{heavymaj2}):
\beq
M_{\nu_R} =
\left (
\begin{array}{ccc}
M & 0 & 0 \\
0 & 0 & M_{4\phi} \\
0 & M_{4\phi} & M_{\phi\phi}
\end{array}
\right )
, \;\;\;\;
V_{\ell_L}^{m \dagger} m_{\nu}^D = \left (
\begin{array}{ccc}
0 &  \alpha s_{\psi} & 0 \\
0 & \alpha c_{\psi} & 0 \\
0 & 0 & \gamma
\end{array}
\right )
\label{simplems}
\eeq
where our approximations are to neglect $M'$ - but not to make
any other {\it a priori} assumption about the relative magnitudes
of entries in $M_{\nu_R}$ - and to neglect terms in 
$V_{\ell_L}^{m \dagger} m_{\nu}^D$  that are
${\cal O}(\bar{\phi}_4)$ - again with no 
{\it a priori} assumption about the relative magnitudes
of other entries. These are parametrized by $\alpha, \gamma$ and
an angle $\psi$, and we denote sin$\psi$ by $s_{\psi}$, etc..
The first approximation could be motivated if
$\phi_3$ is negligible~\cite{ELLN}, and $M$ is eventually
generated by some other effect: as we shall see, the magnitude of
$M$ is not essential for this simplified analysis. On the other
hand, its consistency would require ${\bar F}_5$ to be quite large,
as could occur in the strong-coupling limit of $M$ theory, whilst the
unknown combination $T_2 T_5 \sim {\bar \phi}_4$.

The inputs (\ref{simplems}) yield the following effective light-neutrino
mass matrix
in the weak interaction basis for the neutrinos
\beq
m_{eff} = 
\left (
\begin{array}{ccc}
-{M_{\phi\phi} \alpha^2 s_{\psi}^2 \over M_{4\phi}^2} & -{M_{\phi\phi} s_{\psi} c_{\psi} \over M_{4\phi}^2} &
{\alpha \gamma s_{\psi} \over M_{4\phi}} \\
-{M_{\phi\phi} s_{\psi} c_{\psi} \over M_{4\phi}^2} & -{M_{\phi\phi} c_{\psi}^2 \over M_{4\phi}^2} &
{\alpha \gamma c_{\psi} \over M_{4\phi}} \\
{\alpha \gamma s_{\psi} \over M_{4\phi}} & {\alpha \gamma c_{\psi} \over M_{4\phi}} & 0
\end{array}
\right )
\label{simplemeff}
\eeq
Transforming to the basis $(c_{\psi} \nu_e - s_{\psi} \nu_{\mu}),
(s_{\psi} \nu_e + c_{\psi} \nu_{\mu}), \nu_{\tau}$, $m_{eff}$
in (\ref{simplemeff}) is easily seen to have the form:
\beq
m_{eff} = \left (
\begin{array}{ccc}
0 & 0 & 0 \\
0 & -{M_{\phi\phi} \alpha^2 \over M_{4\phi}^2} & {\alpha \gamma \over M_{4\phi}} \\
0 & {\alpha \gamma \over M_{4\phi}} & 0
\end{array}
\right )
\label{simplermeff}
\eeq
Using the $2 \times 2$ analysis in Section 2, we therefore see
that the three mass eigenstates are:
\bea
\nu_1 \equiv c_{\psi} \nu_e - s_{\psi} \nu_{\mu} & : \;\;\;
& m_1 = 0 \\
\nu_2 \equiv c_{\eta} (s_{\psi} \nu_e + c_{\psi} \nu_{\mu}) - s_{\eta}
\nu_{\tau} & : \;\;\; & m_2 = {2 \gamma^2 \over M_{\phi\phi} + \sqrt{M_{\phi\phi}^2 + 4 M_{4\phi}^2
(\gamma/\alpha)^2}} \\
\nu_3 \equiv s_{\eta} (s_{\psi} \nu_e + c_{\psi} \nu_{\mu}) + c_{\eta}
\nu_{\tau} & : \;\;\; & m_3 = {2 \gamma^2 \over M_{\phi\phi} - \sqrt{M_{\phi\phi}^2 + 4 M_{4\phi}^2
(\gamma/\alpha)^2}}
\label{eigenmasses}
\eea
where
\beq
{\rm sin}^2 2 \eta = { 4 (M_{4\phi} \gamma / \alpha )^2 \over M_{\phi\phi}^2 +  4 (M_{4\phi} \gamma /
\alpha )^2}
\label{defineeta}
\eeq
These simple results equip us to answer some of the questions raised 
by the phenomenological analysis of the data.

We see that one neutrino is massless in this simplified picture, but we
expect it to acquire a small mass when some of the other mixing effects in
(\ref{leptontwo},\ref{twomasses},\ref{heavymaj2}) are taken into account.
The ratio
$|m_3 / m_2|$ may be $\gg 1$ if
$|M_{4\phi} \gamma | \ll |M_{\phi\phi} \alpha|$, or $\approx 1$ if $|M_{4\phi} \gamma | \ll |M_{\phi\phi}
\alpha|$. However, obtaining a large hierarchy $|m_3 / m_2| \sim 10$,
as would be required if $m_3 \sim 10^{-3/2}$~eV and $m_2 \sim
10^{-5/2}$~eV,
seems to require less fine tuning than
obtaining near-degeneracy:  $(m_3^2 - m_2^2) / m_3^2 \sim 1/100$, as
would be required if the neutrino masses were to be cosmologically
significant: $m_{2,3} \sim 1$~eV. Moreover, any such degeneracy
would be very sensitive to higher-order corrections, and
there is no apparent
mechanism for making $\nu_1$ approximately degenerate with $\nu_{2,3}$,
as would also be required in this scenario.

Large mixing appears naturally in the $\nu_{\mu} - \nu_e$ sector
for generic values of $\psi$,
but its magnitude is model-dependent. In particular, there is the
logical
possibility of a cancellation between the mixing 
in $(V^m_L)^{\dagger}$ and
$m_{\nu}^D$ that could suppress it significantly: sin$\psi \ll
1$. Nevertheless, the large-angle MSW solution seems quite plausible.
Large mixing in the $\nu_{\mu} - \nu_{\tau}$ sector is also quite
generic. The simplified parametrization above might indicate an
apparent conflict with a large hierarchy: $|m_3 / m_2| \gg 1$. However,
following the discussion in Section 3, we expect large mixing and
a large hierarchy to be quite compatible when the full
parameter space of (\ref{leptontwo},\ref{twomasses},\ref{heavymaj2}) is
explored.
Moreover, we should also remember
that the effective neutrino mixing angle may be amplified by
renormalisation
group effects in the case of large tan${\beta}$, as discussed in
Section 4 and
seen in Fig. 5, so we need not require that the maximal mixing
be present already at the GUT scale.

We now consider the complementary possibility, where
the field $\phi_3$ develops a large
vev. The larger is $\phi_3$, the
smaller are $m_{\nu_\mu}$ and $m_{\nu_e}$
with respect to $m_{\nu_\tau}$. At this stage, we assume for
simplicity that $\phi_3\approx 1$
and we define coefficients that keep track of
the relation between the various entries
of $M_{\nu_R}$.
Then, we write $M_{\nu_R}$ in
(\ref{hema}) as
\begin{equation} 
M_{\nu_R} = 
\left (
\begin{array}{ccc}
\bar{F}_5 \bar{F}_5  \bar{\phi}_4  &
\bar{F}_5 \bar{F}_5  \Delta_2 \Delta_5  & 0 \\
\bar{F}_5 \bar{F}_5  \Delta_2 \Delta_5  &
0 & \bar{F}_5  \bar{\phi}_4  \\
0 & \bar{F}_5  \bar{\phi}_4  
& \Delta_2 \Delta_5 \bar{\Phi}_{23} T_2 T_5
\end{array}
\right) \equiv
\left (
\begin{array}{ccc}
f y^2 & x y^2 & 0 \\
xy^2 & 0 & f y \\
0 & f y & t x
\end{array}
\right) 
\end{equation}
where
$\Delta_2 \Delta_5 \equiv x, T_2 T_5 \equiv t$,
$\bar{\phi}_4 \equiv f$
and $\bar{F}_5 \equiv y$.
For the Dirac mass matrix, as in the previous case,
we have the possibility of cancellations between
the charged lepton and neutrino mixing matrices.
To simplify the presentation in
terms of the mass matrices, we describe two cases separately.

In the absence of a cancellation, the 
Dirac mass matrix in the weak-eigenstate basis
is of the form
\begin{equation} 
V_{\ell_L}^{m\dagger}.m_{\nu}^D \approx
\left (
\begin{array}{ccc}
1 & -gx & 0 \\
gx  &  1 & 0 \\
0 & 0 & 1
\end{array}
\right )
\left (
\begin{array}{ccc}
f x & 1 & 0 \\
f & x & 0 \\
0 & 0 & y
\end{array}
\right )
\end{equation}
where we have dropped terms of 
order $x^2$ in $(V_{\ell_L})^{m \dagger}$.
Then,
\begin{equation}
m_{eff} \propto
\left (
\begin{array}{ccc}
-f^4 (-1+g^2)^2 x^2 
&  f^4 (-1+g) x (1+gx^2) 
&  -f^2(1+ (-2+g)x^2)y^2 \\
%%%%% 2nd part %%%%%%
\hspace*{0.8 cm}
 + ftx (1+...)  & 
\hspace*{0.8 cm}
 + ftx^2 (1+...) &
 \\ & & \\
 f^4 (-1+g) x (1+gx^2)  &
-f^4(1+gx^2)^2 & 
f^2 x (1+ g(-1+2x^2))y^2  \\
\hspace*{0.8 cm}
 + ftx^2 (1+...) &
-ftx^3 (1+...)   & \\
& & \\
-f^2(1+ (-2+g)x^2)y^2 
& f^2 x (1+ g(-1+2x^2))y^2  
& -4 x^2 y^4
\end{array}
\right )
\end{equation}
while
\begin{equation}
\sin^2 2 \theta_{23} = \frac{ 4 f^4 (x-gx+2gx^3)^2 y^4
}{
(4 f^4 (x-gx+2gx^3)^2 y^4)+
( f^4(1+gx^2)^2 + ftx^3(3+...) -4 x^2 y^4)
}
\end{equation}
We see therefore that if
$\bar{\phi}_4 \approx \bar{F}_5,F_1$,
as would be expected in weak-coupling
unification schemes,
the entries of $m_{eff}$ are all of the same order of magnitude.
In this case, as we discussed in the
previous
phenomenological analysis, large 
$\nu_{\mu}-\nu_{e}$ and 
$\nu_{\mu}-\nu_{\tau}$ mixings are both generated,
whilst cancellations between the various terms can
lead to large hierarchies between the neutrino
masses.

Suppose now that a cancellation between the charged
lepton and the neutrino mixing matrices takes place.
In this case, we write 
\begin{equation} 
V_{\ell_L}^m  = 
\left (
\begin{array}{ccc}
1 & g x & 0 \\
-g x & 1 & 0 \\
0 & 0 & 1
\end{array}
\right )
\end{equation}
where $g \approx (1-a \bar{\phi}_4)/x^2$:
this leads to (1,2) and (2,1)
entries in the Dirac mass matrix
of the order of
$\bar{\phi}_4$.
In this case, 
\begin{equation}
m_{eff} \propto
\left (
\begin{array}{ccc}
-2a f^2 t x (-1 + x^2) & -f t (1+x^2) (-1+x^2) & f^2 (-1+x^2) y^2 \\
-f t (1+x^2) (-1+x^2) & f t (1-2x^2)/x & f^2 (-1+x^2)y^2 /x \\
f^2 (-1+x^2) y^2 & f^2 (-1+x^2)y^2/x & -x^2 y^4
\end{array}
\right )
\end{equation}
and we see a difference from the
previous example, in that now
all the entries of the (1,2) sector are
multiplied by $t$, and therefore may be suppressed
if $T_2 T_5$ is small. The entries for the
(2,3) sector are similar to the previous case,
with the modification that the (2,2) entry can be very small. 
Large (2,3) mixing is again generated for
$\bar{\phi}_4 \approx \bar{F}_5,F_1$.

We conclude this Section by commenting on the possible order
of magnitude of neutrino masses in this model, using (\ref{eigenmasses})
as our guide. The factor $\gamma$ appearing in the numerator 
and denominator is
expected to be ${\cal O}(1) \times M_W$, since it comes from a
third-order coupling. The same estimate applies to the
factor $\alpha$ appearing in part of the denominator. 
The factors $M_{4\phi},M_{\phi\phi}$
that also appear there originate from seventh-order couplings,
and hence are expected to be considerably smaller, with a typical
estimate being ${\cal O}(10^{-4 \pm 1}) \times M_s$. Taking
$M_s \sim 10^{16}$ to $10^{18}$~GeV, we might guess that
$M_{\phi\phi}, M_{4\phi} \sim 10^{13 \pm 2}$~GeV. 
Our final estimate is therefore that
\beq
m_3 \sim 10^{0 \pm 2}\; {\rm eV}
\label{estimate}
\eeq
which is consistent (within our uncertainties) with the indication
provided by the super-Kamiokande data~\cite{SKam} that $m_3^2 \geq 
10^{-3}$~eV$^2$.

We conclude that the flipped $SU(5)$ model appears capable, within
its considerable uncertainties, of proving to be consistent with
the magnitudes of the neutrino masses and mixing angles suggested
by experiment.

\section{Conclusions}

In this paper we have first analyzed possible patterns
of neutrino masses and mixing compatible with the
atmospheric and solar neutrino deficits from a purely
phenomenological point of view. In particular, we have
emphasized that large neutrino mixing as suggested by the
super-Kamiokande atmospheric neutrino data~\cite{SKam} does not
necessarily require near-degeneracy between a pair of
neutrino masses. We have discussed possible patterns of
$2 \times 2$ and $3 \times 3$ Dirac and massive Majorana mass matrices
that are compatible these and MSW interpretations of the solar neutrino
data. We have also provided semi-analytic formulae for
renormalization-group effects, and re-evaluated their
impact on the light-neutrino mixing angles, which may well be important.
Equipped with this phenomenological
background, we have gone on to discuss neutrino masses and mixing
in general models with a $U(1)$ flavour symmetry, and in
a `realistic' flipped $SU(5) \times U(1)$ model derived from string.

The discussion of this part of our paper serves to
reinforce the message that, whilst
the string selection
rules restrict the forms of terms that one may obtain from
a specific string-derived
model, it is nevertheless possible to obtain 
realistic patterns of fermion masses and mixings.
We had demonstrated this previously for quarks and charged
leptons, and have extended that discussion to neutrinos
in this paper. In particular, we have shown that
it is possible to have contributions which lead to
plausible hierarchical magnitudes of neutrino masses,
a large mixing angle that could explain the 
atmospheric neutrino deficit, and either the
large- or the small-angle MSW solution to the
solar neutrino deficit. 

The higher-dimensional operators that we obtain depend only on the choice of
string model, but the detailed forms of the mass matrices clearly
depend on the choice of flat direction. This introduces some
ambiguity, and work remains to be done to demonstrate that the
choice made in this paper remains valid to higher orders in the
effective superpotential derived from the string model.
Despite this apparent freedom in the choice of vev's, the room
for manoeuvre in such a string-derived model is quite
restricted, and we find it interesting that
it is nevertheless possible to obtain a
realistic scheme for fermion masses and mixings
and even obtain solutions with 
large neutrino oscillations.

We conclude by stressing again
some aspects of our specific model analysis that
might be of general interest to model-builders.
(i) Once outside the framework of $SO(10)$-like
models, there is no general expectation that the neutrino
Dirac mass matrix should be equivalent to the $u$-quark mass matrix,
in particular because (ii) charged-lepton mixing may also be
significant, and different from that of $d$-type quarks.
Moreover, (iii) mixing in the heavy Majorana
mass matrix is in general mismatched relative to the other mass
matrices, leading to a generic expectation of large
mixing angles for the light neutrinos. Specifically,
this can occur because (iv) the effective $\nu_R$ states may include
gauge-singlet fields that are not related by GUT symmetries
to any Standard Model particles. Finally, we note that,
because the heavy Majorana mass matrix elements typically arise
from higher-order non-renormalizable terms, (v) it is quite natural
that the mass eigenvalues be much smaller than $M_s$ or $M_{GUT}$,
possibly with values ${\cal O}(10^{13})$~GeV, 
as would be required to generate
a light neutrino mass ${\cal O}(0.1)$~eV.
 
{\bf Acknowledgements}

The work of D.V.N. has been supported in part by the U.S. Department of
Energy under grant DE-FG03-95-ER-40917.

\normalsize

\vspace*{0.3 cm}

\begin{center}
{\bf Appendix}
\end{center}
In this appendix we tabulate for completeness
the field assignment of the `realistic'
flipped $SU(5)$ string model~\cite{aehn},
as well as the basic conditions used in~\cite{ELLN} to obtain consistent
flatness conditions and acceptable Higgs masses.

\begin{table}[h]
\centering
\begin{tabular}{|c||c||c|}
\hline
$F_1(10,\frac{1}{2},-\frac{1}{2},0,0,0)$ &
$\bar{f}_1(\bar{5},-\frac{3}{2},-\frac{1}{2},0,0,0)$ &
$\ell_1^c(1,\frac{5}{2},-\frac{1}{2},0,0,0)$ \\
 
$F_2(10,\frac{1}{2},0,-\frac{1}{2},0,0)$ &
$\bar{f}_2(\bar{5},-\frac{3}{2},0,-\frac{1}{2},0,0)$ &
$\ell_2^c(1,\frac{5}{2},0,-\frac{1}{2},0,0)$ \\
 
$F_3(10,\frac{1}{2},0,0,\frac{1}{2},-\frac{1}{2})$ &
$\bar{f}_3(\bar{5},-\frac{3}{2},0,0,\frac{1}{2},\frac{1}{2})$ &
$\ell_3^c(1,\frac{5}{2},0,0,\frac{1}{2},\frac{1}{2})$ \\
 
$F_4(10,\frac{1}{2},-\frac{1}{2},0,0,0)$ &
$f_4(5,\frac{3}{2},\frac{1}{2},0,0,0)$ &
$\bar\ell_4^c(1,-\frac{5}{2},\frac{1}{2},0,0,0)$ \\
 
$\bar{F}_5(\overline{10},-\frac{1}{2},0,\frac{1}{2},0,0)$ &
$\bar{f}_5(\bar{5},-\frac{3}{2},0,-\frac{1}{2},0,0)$ &
$\ell_5^c(1,\frac{5}{2},0,-\frac{1}{2},0,0)$ \\
\hline
\end{tabular}
%\caption
\label{table:4}
 
\vspace*{0.5 cm}
 
\centering
\begin{tabular}{|c||c||c|}
\hline
$h_1(5,-1,1,0,0,0)$ & $h_2(5,-1,0,1,0,0)$ & $h_3(5,-1,0,0,1,0)$ \\
$h_{45}(5,-1,-\frac{1}{2},-\frac{1}{2},0,0)$ &  \\
\hline
\end{tabular}

\vspace*{0.5 cm}
 
\centering
\begin{tabular}{|c||c||c|}
\hline
$\phi_{45}(1,0,\frac{1}{2},\frac{1}{2},1,0) $ &
$\phi_{+}(1,0,\frac{1}{2},-\frac{1}{2},0,1) $ &
$\phi_{-}(1,0,\frac{1}{2},-\frac{1}{2},0,-1) $ \\
$\Phi_{23}(1,0,0,-1,1,0)$ &
$\Phi_{31}(1,0,1,0,-1,0)$  &
$\Phi_{12}(1,0,-1,1,0,0)$ \\
$\phi_i(1,0,\frac{1}{2},
-\frac{1}{2},0,0)$ &
$\Phi_i(1,0,0,0,0,0)$ & \\
\hline
\end{tabular}
 
\vspace*{0.5 cm}
 
\centering
\begin{tabular}{|c||c||c|}
\hline
$\Delta_1(0,1,6,0,-\frac{1}{2},\frac{1}{2},0)$ &
$\Delta_2(0,1,6,-\frac{1}{2},0,\frac{1}{2},0)$ &
$\Delta_3(0,1,6,-\frac{1}{2},-\frac{1}{2},0,
\frac{1}{2})$ \\
$\Delta_4(0,1,6,0,-\frac{1}{2},\frac{1}{2},0)$ &
$\Delta_5(0,1,6,\frac{1}{2},0,-\frac{1}{2},0)$ & \\
\hline
\end{tabular}

\vspace*{0.5 cm}
 
\centering
\begin{tabular}{|c||c||c|}
\hline
$T_1(0,10,1,0,-\frac{1}{2},\frac{1}{2},0)$ &
$T_2(0,10,1,-\frac{1}{2},0,\frac{1}{2},0)$ &
$T_3(0,10,1,-\frac{1}{2},-\frac{1}{2},0,\frac{1}{2})$ \\
$T_4(0,10,1,0,\frac{1}{2},-\frac{1}{2},0)$ &
$T_5(0,10,1,-\frac{1}{2},0,\frac{1}{2},0)$ & \\
\hline
\end{tabular}

\vspace*{0.2 cm}

{\small Table 3: 
{\it The chiral superfields are listed with their
quantum numbers \cite{aehn}.
The $F_i$, $\bar{f}_i$, $\ell_i^c$,
as well as the
$h_i$, $h_{ij}$ fields and the singlets
are listed with their
$ SU(5) \times U(1)' \times U(1)^4$ 
quantum numbers. 
Conjugate fields have opposite $U(1)' \times U(1)^4$
quantum numbers.
The fields $\Delta_i$ and $T_i$ are tabulated in terms
of their $U(1)' \times SO(10) \times SO(6) \times U(1)^4$
quantum numbers. }
}
\end{table}
 
As can be seen, the matter and
Higgs fields in this string model carry additional charges under
additional $U(1)$ symmetries~\cite{aehn}. There exist various
singlet fields, and  hidden-sector 
matter fields which transform
non-trivially under the $SU(4)\times SO(10)$ gauge symmetry,
some as sextets under $SU(4)$, namely $\Delta_{1,2,3,4,5}$, and some as
decuplets under $SO(10)$, namely $T_{1,2,3,4,5}$. There are also
quadruplets of the $SU(4)$ hidden symmetry which possess fractional
charges. However,  these are confined and
will not concern us further.

The usual flavour assignments of the light
Standard Model particles in this model are as follows:
\bea
\bar{f}_1 : \bar{u}, \tau, \; \; \;
\bar{f}_2 : \bar{c}, e/ \mu, \; \; \;
\bar{f}_5 : \bar{t}, \mu / e \nonumber \\
F_2 : Q_2, \bar{s}, \; \; \;
F_3 : Q_1, \bar{d}, \; \; \;
F_4 : Q_3, \bar{b} \nonumber \\
\ell^c_1 : \bar{\tau}, \; \; \;
\ell^c_2 : \bar{e}, \; \; \;
\ell^c_5 : \bar{\mu}
\label{assignments}
\eea
up to mixing effects, which are discussed in more detail in Section 7.
We chose non-zero vacuum expectation values for the 
following singlet and hidden-sector fields:
\beq
\Phi_{31}, \bar{\Phi}_{31}, \Phi_{23}, 
\bar\Phi_{23},\phi_2, \bar{\phi}_{3,4},  \phi^-, 
 \bar\phi^+ ,  \phi_{45},  \bar{\phi}_{45}, \Delta_{2,3,5}, T_{2,4,5}
\label{nzv}
\eeq
The vacuum expectation values of the hidden-sector fields
must satisfy the additional constraints 
\beq
 T_{3,4,5}^2=T_i\cdot T_4=0,\,\, \Delta_{3,5}^2=0,\,\, T_2^2+\Delta_2^2=0
\label{hcon}
\eeq
For further discussion, see~\cite{ELLN} and references therein.

%%%%%%%%%%%%

\end{document}